# Understanding the online behavior and risks of children: results of a large-scale national survey on 10-18 year olds

**Evangelia Daskalaki, Katerina Psaroudaki, Marieva Karkanaki, Paraskevi Fragopoulou**[*]

*Institute of Computer Science*
*Foundation for Research and Technology - Hellas (FORTH)*
*N. Plastira 100, Vassilika Vouton, GR-70013 Heraklion, Crete, Greece*

**ABSTRACT**

The internet has opened up new horizons of knowledge, communication and entertainment in our lives. Through this, young people are presented with a wealth of opportunities and activities that can enhance their skills and empower their knowledge and creativity. However, the online engagement of young people often comes with significant risks, encountered by children accidentally or deliberately. The emergence of new online services at an unprecedented speed and innovation brings the need, internationally, for a constant monitoring and investigation of the rapidly changing landscape and the associated emerging risk factors that could potentially jeopardize children's development, opportunities, and lives.

The Greek Safer Internet Center (Daskalaki, Psaroudaki, & Fragopoulou, 2018) conducted two large-scale surveys to understand children's internet engagement, aiming to contribute towards improved child protection policies that could guide the efforts of key stakeholders towards a safer cyberspace. Furthermore, the reported results may also lead policy implication that could potentially benefit the development of effective awareness tools, parental controls, safety regulations, etc.

The first survey took place at the end of 2018, with the approval of the Greek Ministry of Education and Religious Affairs, and was conducted online among 14,000 pupils aged 10-18 years from 400 schools spread in five different urban areas of Greece. A follow up survey was realized the following year, among 13,000 students of the same age group from 500 school units in six different prefectures of Greece. To our knowledge, it is the first time national surveys of such scale are conducted in Greece under the approval of the Ministry of Education.

This paper presents the analysis of the collected data, and describes the underlined methodology based on which the survey was formulated and conducted according to international standards. The surveys were drafted around a number of specific thematic areas, namely *internet use and online behavior*, *parental engagement, confidence level of children, digital literacy, social media,* and *online risks*. The results were analyzed based on *educational level* and *gender*, the two variables that emerged as the most significant differentiating factors, while parameters such as *sampling area* and *parents' educational level* appeared beyond statistical significance.

The survey showed that almost all children use the internet daily, while nine out of ten own a mobile device, mainly mobile phone, fact that renders internet access the ultimate personal experience. The majority of children start to use the internet at the age of 7-8, an age constantly decreasing over time, while one in five children begins to use the internet at the very young and sensitive age of 4-6 years. Girls are mainly engaged in social networks and boys in online gaming. Furthermore, children make use of social networks at an increasingly younger and not permissible age, often creating profiles without the consent of their parents. A significant percentage of children put themselves at risk by adopting wrong practices such as accepting friends' requests from strangers, meeting people they got to know online, posting very personal photos on the web, and in general sharing material without thinking of the possible consequences. An important point emerging from the answers is the fact that one in five children admits to have become the target of online harassment at some point in its 'digital life'. Concerning the key issue of excessive use of the internet, almost half the children admit to neglect their activities for the internet, while about one out of three children believes to make excessive use of the internet. Yet, from the self-assessment section, it emerges that the largest percentage of children feel confident that they know how to use the internet safely and to cope with emerging risks. Another main outcome of the survey was that most parents seem to ignore the importance of setting limits and rules concerning internet use to their kids from a very young age, and they lack the required vigilance on the topic.

---

[*] Corresponding author: Paraskevi Fragopoulou (fragopou@ics.forth.gr), Foundation for Research and Technology-Hellas (FORTH), Institute of Computer Science, N. Plastira 100, Vassilika Vouton, GR-70013 Heraklion, Crete, Greece

# 1. Introduction

The internet has opened up new horizons of knowledge, communication and entertainment in our lives. Through the internet, young people have a wealth of opportunities for activities that can enhance their skills and empower their knowledge and creativity. Nowadays, more than ever, children gain online access at a very early age with next generation personal devices through rapid mobile networks to rich media sites (Directorate-General for Communication, 2019). At the same time, industrial partners race to offer new compelling services to highly engage young users, schools seek to maximize online opportunities and creative engagement through media literacy skills, while parents', carers' and the society as a whole tries to cope with the potentially emerging risks.

The internet knows no limits and accepts no centralized control, and although this openness translates into an unprecedented network of unique opportunities, it comes at a price. The online engagement of youth people often presents significant risks, either we talk about their first steps in internet use, or about more experienced users, risks have been identified and are encountered by children every day, accidentally or sometimes deliberately due to risky and hazardous activity. Online harassment, addiction, cyberbullying, theft of personal data, sexting, misinformation are just some of the challenges today's parents and teachers have to face in order to ensure a safe and quality 'digital life' for children. *The emergence of new online services at unprecedented speed and innovation brings the need, internationally, for a constant monitoring and investigation of the rapidly changing landscape and the associated emerging risk factors that could potentially endanger children's development, opportunities, and lives.*

The Greek Safer Internet Awareness Center SaferInternet4Kids (Greek Safer Internet Awareness Center, 2019) started its operations in June 2016 under the aegis of the Institute of Computer Science of the Foundation for Research and Technology - Hellas (FORTH, 2019). Together with SafeLine (Greek Hotline for illegal Internet content, 2019) and the Help-Line (Help-Line, 2019), they constitute the Greek Safer Internet Centre (SIC), member of the Insafe/INHOPE networks (Insafe, 2019). The Greek SIC operates in the framework of the SI4Kids CET-Telecom project with the support of the European Commission, and aims to monitor and safeguard the online safety of children. Similar centers operate, also with the support of the European Commission, in almost all European countries. The Greek SIC, was recently nominated by the Greek General Secretariat for Telecommunications and Post to act as a national representative in the [European Commission's Safer Internet for Children Expert Group](). To this end, the Greek SIC conducts periodic surveys about the online behavior of children and adolescences, with the aim to contribute towards improved child protection policies that could potentially guide the steps the country needs to take to advocate for a safer cyberspace.

This paper provides a thorough analysis of the latest survey results concerning the online behavior of children, designed to collect detailed information from youth on a wide range of internet experiences and online risks including harassment, exposure to inappropriate content, and self-produced sexual images. Furthermore, sections on social media use and excessive use as well as internet safety education and risk prevention efforts are included.

There has been a wide range of national surveys on the topic of online behavior of youth (Livingstone, Davidson, Bryce, Batool, & Nandi, 2017), (Park & Kwon, 2018), (Mitchell & Jones, 2012), (Madariaga, et al., 2017) (Global Kids Online, 2017) (Smahel, et al., 2020) (Livingstone S. , Developing social media literacy: How children learn to interpret risky opportunities on social network sites, 2014) (Livingstone, Mascheroni, & Staksrud, European research on children's internet use: assessing the past and anticipating the future, 2018). This paper presents the first large-scale surveys conducted on this topic in Greece under the approval of the Greek Ministry of Education and Religious Affairs, among 14,000 pupils aged 10-18 years old, with a follow up the next year on 13,000 students. The findings of these surveys are compared against a literature review of other similar surveys conducted by Livingston et al. 2017 (Livingstone, Davidson, Bryce, Batool, & Nandi, 2017) as well as EU Kids Online (Smahel, et al., 2020) .

As these are the largest sample surveys ever on this topic at national level in Greece, they can also serve as a valuable tool for policy planning by decision makers in order to inform and educate students on the safe use of the internet. Furthermore, it could serve as a useful tool for the educational community, as research reveals children's risky habits when using the internet, which can result to difficult or even dangerous situations. As stated by Prof. Livingston et.al., "*It is well known that risk/vulnerability factors are likely to shape children's online experiences, and this is mediated by the ways in which children develop emotionally, cognitively, social relationships and need for support,* *and their peer cultures; however, it remains difficult to know the moment when children give in to specific online risks."* (Livingstone, Mascheroni, & Staksrud, European research on children's internet use: assessing the past and anticipating the future, 2018). Thus, it is essential to conduct surveys on this topic, in order to be better prepared nationwide and to put focused effort designing targeted educational strategies.

The target survey population were children 10-18 years old. The first survey took place from November to December 2018 on a random sample of students from 400 public schools (Elementary schools, Gymnasiums and Lyceums) from various urban areas spread around the country. The total number of participants were approximately 14,000 students. A second survey, took place the following school year, from November to December 2019, on a random sample of students from 500 public school (Elementary schools, Gymnasiums and Lyceums) again from several urban locations in the country (not necessarily the same). The total number of participants in the second survey were approximately 13,000 students.

The selected locations included islands and regions in proximity to the borderline of Greece (Fig 1). Data were collected anonymously, with the method of Computer-Assisted Web Interviewing (CAWI), via online questionnaires. Questionnaires were filled online individually by students during their IT course, and under the discrete supervision, but not interference, of their instructor.

The variables on which the analysis of the results was based were education level (thus age group, as there is direct correlation) and gender. The remaining of the variables recorded, such as location of the sample unit and educational level of parents, did not demonstrate any statistical difference in the result analysis, and thus presenting them and examining them further, was not deemed necessary. Consequently, the analysis of the results is performed based on demographic factors, rather than social and economic ones.

The main focus of the survey and the main observations drawn can be summarized as follows:

- The survey investigates children's first steps online, such as age of first use, first device used to go online, age first smart phone acquired, etc. Internet use and online behavior of children are also examined, such as type of online engagement, main device used for online access, frequency of use and time spend, activities that monopolize children's time online based on age and gender, etc.
- An issue that attracted attention is the reaction of children when confronted with an emerging risk, such as where to turn for help or their level of knowledge on how to report.
- Furthermore, the confidence level of children when online was investigated, with the intention to understand how this is perceived by them, independently of whether it correspond to a real or plasmatic impression. As emerges from the self-assessment, the largest percentage of children feel confident that they know how to use the internet safely and how to cope when confronted with emerging risks.
- The survey helped emerge online risks, such as harassment and cyberbullying, accepting friends' requests from strangers, meeting with strangers, exposure to inappropriate content, sending of intimate information/photos, etc. An important point emerging from the answers is the fact that one out of five children admits to have become the target of online harassment at some point in its 'digital life'. A significant percentage of children put themselves at risk by adopting wrong practices such as accepting friendship requests from strangers, meeting people they got to know online, sharing very personal photos on the web, sharing material without thinking of the possible consequences.
- Concerning the key issue of excessive use of the internet, almost half the children admit to neglect their daily activities in order to be online, while about one out of three children either admits or suspects to have an internet addiction problem.
- Regarding social media, an analysis of the results shows engagement of children at a decreasing, not yet permissible age, most often without the consent of their parents.
- Parental engagement and mediation appears to be inadequate, as involvement of parents, especially those of older children, is minimal and rules, regulations and limits seldom applied, highlighting the need for improved parental supervision and possibly improved awareness raising and parental understanding.

- However, results show that parents of younger children, the new generation of parents that also grew up with technology, is actively involved in their child's first steps in the digital world. On the contrary, parents of older children are not sufficiently engaged to safeguard their safety, ignore the importance and fail to impose adequate rules and set limits, and they lack the required vigilance on the topic.
- As shown by the results, primarily parents/caregivers and, secondarily, older siblings undertake the task of guiding children in their first steps online. Very little help is recorder to come from educators, demonstrating the need for enhanced involvement of schools in internet safety education, from a very early age.
- These observations, in combination with the earlier occupation of children with the online world, dictates the need for a more organized treatment of the topic by the educational system, starting education in schools in an organized way, incorporating material into school curricula, from as early as kinder garden and even preschool period. An alarming observation is drown from the fact that school appears to be non-existent in coping with emergency/anxiety online occurrences, explained by the fact that schools in Greece lack, in general, the required structure to deal with emergency situations.

The remaining of the paper is subdivided into the following sections: Section "Methodology" gives the detailed methodology of the survey as well as ethical aspects and limitations, followed by section "Internet Use", which outlines the general attitude of children concerning online behavior. Section "Online Risks", is devoted to online risks and the extent to which children can identify them properly and protect themselves. Section "Future Work and Conclusions" sums up the conclusions of the paper and highlights the next steps of the Greek Safer internet Center.

## 2. Methodology

### 2.1 Sample - Data collection

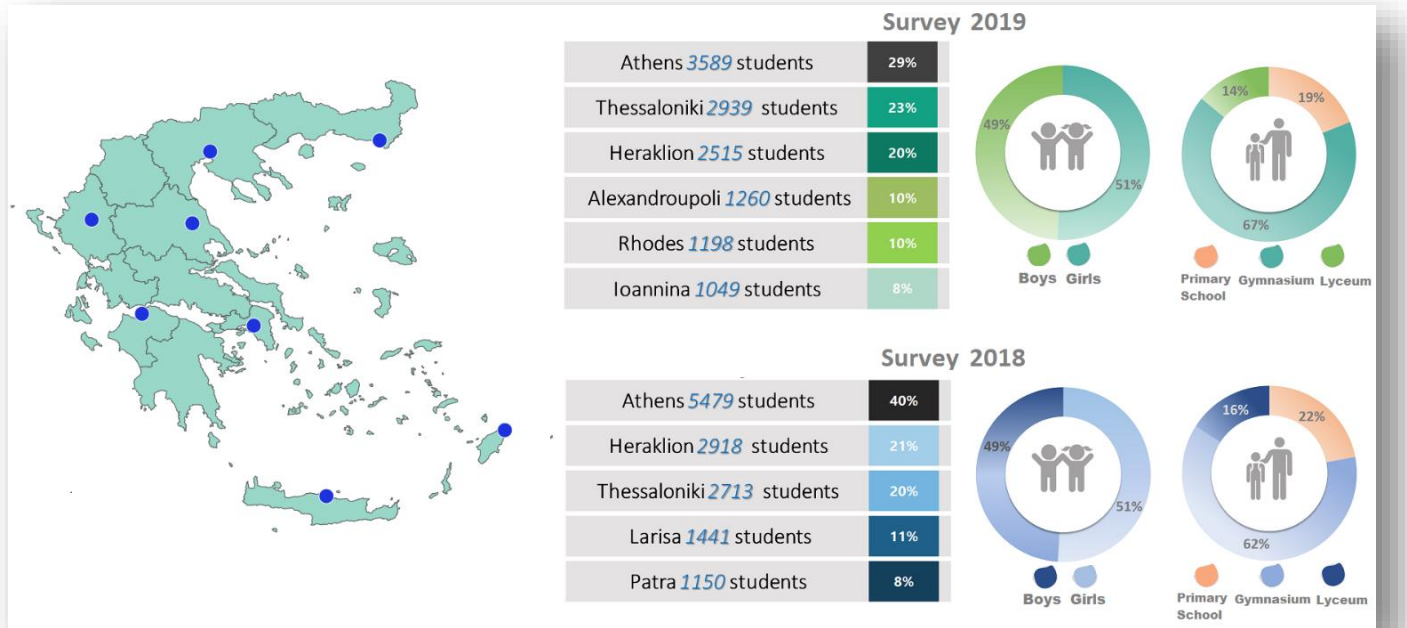

**Fig 1:** General Survey Data

The main survey took place in the period November-December 2018 under the approval of the Greek Ministry of Education and Religious Affairs (approval number for Elementary schools F15/109951/173050/D1 and for Secondary schools 178261/D2). The unit of analysis was fifth and sixth grade Elementary school students (10-12 year olds), Gymnasium students (equivalent to Middle/Junior High schools, 12-15 year olds) and Lyceum students (equivalent to Senior High school, 15-18 year olds). More specifically, a random sample of approximately 14,000 pupils aged 10-18 years from 400 schools in Athens, Thessaloniki, Heraklion-Crete, Larissa, and Patra, each one belonging to a different regional municipality of Greece, took part in the survey.

A second online survey was realized the following year, again with the approval of the Greek Ministry of Education and Religious Affairs (approval number F15/152679/167587/D1) in the period November-December 2019 among approximately 13,000 students aged 10-18 years old from 500 school units in Athens, Thessaloniki, Alexandroupoli, Ioannina, Heraklion-Crete, and Rhodes-Dodecanese. Three of the locations were common in both surveys. The locations and distribution of the samples are shown in Fig 1.

All selected locations were urban centers. Athens is the capital and largest in size city of Greece, Thessaloniki is the second largest city. Alexandroupoli is located at the borders line of Greece, in the prefecture of Evros, and Heraklion and Rodos in Aegean islands. The border and islandic regions involved were also major urban areas.

The sample extraction was based on the judging sampling technique (Deming, 1990). The selection of the analysis units was performed by researchers of the Greek Safer Internet Center of the Foundation for Research and Technology-Hellas (FORTH, 2019), Institute of Computer Science, in collaboration with the Greek Ministry of Education to include a representative sample. The survey was announced by the Greek Ministry of Education with an official document sent to the Principals of the involved school units,

Both surveys were conducted in the form of anonymous online questionnaires, which were filled individually and submitted by students using school computers, privately, and not under the observation of their peers. The questionnaire was filled during the IT course, and under the presence and discrete supervision - but not interference

- of the instructor. Last, but not least, the website of the Greek Safer Internet Center uses secure connection (HTTPS) with encryption certificate, to ensure secure communication, protection of the privacy and integrity of the exchanged data while filling out the questionnaire.

From the respondents, 51% were girls and 49% boys in both surveys. Additionally in the 2018 survey, 19% were in Elementary school, 67% in Gymnasium and 14% in Lyceum, while in 2019 these percentages were 22%, 62% and 16%, respectively. The parents/guardians of 56% of the children who took part in the survey had completed Secondary education, while 44% of them had a University degree.

The main demographic variables of the survey are the educational level of the analysis units (Elementary school - $5^{th}$ and $6^{th}$ Grades / Gymnasium / Lyceum). The result collection and analysis were based on education level, rather than age, since these two parameters are directly correlated in the vast majority of cases, as follows:

- **Elementary school:** 10-12 year olds (last two grades of elementary school, $5^{th}$ and $6^{th}$)
- **High school:** 12-15 year olds (equivalent to Middle/Junior High school)
- **Lyceum:** 15-18 year olds (equivalent to Senior High school)

Gender (boy/girl) emerged as a predominant parameter in relation to internet use, and was thus included in the result analysis. The educational level of parents (Basic Education/University Education), a parameter related to the socioeconomic status of the household was also recorded. However, no significant statistical difference emerged, and consequently, it is not included as a feature in this analysis of the results. A similar observation was made for the location of the sampling unit (Athens/ Thessaloniki/ Heraklion/ Larissa/ Patra and Athens/ Thessaloniki/ Alexandroupoli/ Ioannina/ Heraklion/ Rhodes), as the surveys were conducted in urban areas of the country only (border and islandic regions involved were also urban regions). This is most probably the main reason behind the fact that location did not emerge as a differentiating parameter in the analysis of the results. Rural areas, more isolated border and islandic regions, and smaller communities in general, which are in abundance in Greece due to the country's morphology, were not part of the surveys, as they will be the main subject of a new survey. To summarize, the analysis of the results were performed taking into consideration demographic factors (i.e., educational level, gender), rather than social factors (i.e., educational level of parents, economic status).

The main goal achieved by the Greek Safer Internet Center, was to capture children's online habits and to draw safe conclusions that could be used as guidelines for designing awareness strategies for educating children, parents and teachers. The aim is also to include the research results in the more general Cohort study (Song & Chung, 2010), providing insight on the way online habits and perceptions of children and adolescent in Greece change over time.

## 2.2 Questionnaire

The online questionnaire was formulated according to international standards and was released with specific guidelines on how to be filled. Furthermore, it was designed to help derive safe conclusions on how online use evolves over time, in order to draw recommendations for various stakeholders/target groups such as:

- **Children:** Concerning their level of awareness and digital literacy in relation to internet safety and their self-confidence, as they perceive it.
- **Parents:** Concerning the need for improved parental supervision (provide guidance, apply rules, set limits, etc.) and possibly improved awareness raising and parental understanding. Compare the level of involvement for parents of younger children to those of elder ones. Level of parental engagement, especially in first steps of children's online activity, how this evolved in new generations of parents that grew up with digital technology.
- **Schools:** Concerning the degree of guidance provided by educators and the level of readiness in confronting with emerging risks. Involvement of schools in digital safety education (cultivation of vigilance, assistance in resilience development), and digital literacy acquired by existing curricula.
- **State/Ministry of Education:** Concerning educational strategies to be applied involving schools more constructively into the digital education processes.

- **Industry:** Concerning its involvement mostly in relation to social media use and the major issue of permissible age, as well as online games and the development of best practices for the provided services.

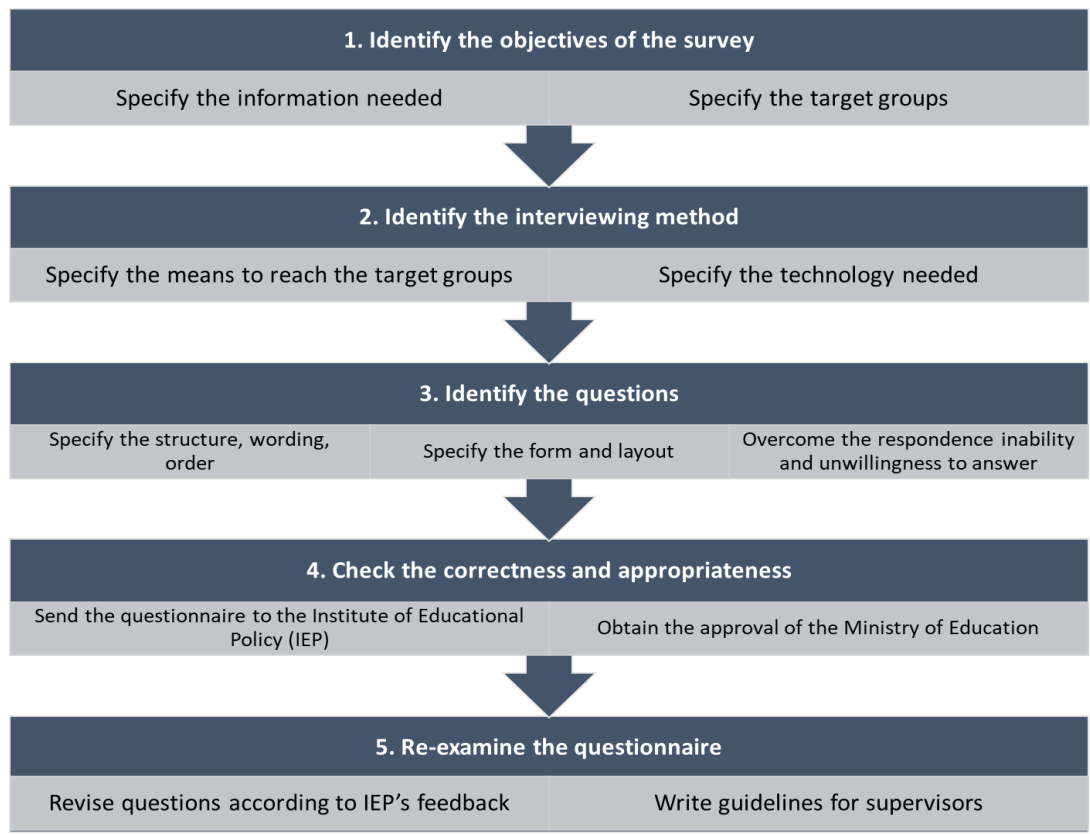

**Fig 2:** Questionnaire Process

The process we followed to design the questionnaire is depicted in Fig 2. First, we identified the objectives of the survey, with the aim to answer specific questions and to compare results across age groups and gender. For this step, we specified the information needed evolving around a number of specific thematic areas, intended to shed light to the following issues:

- **Internet use and behavior:** First online engagement, at what age, ease of access, type of access based on device used (i.e., mobile, personal), type of favorite activity, frequency of access/screen-time, guidance received, how children's online experiences change over time, changes in the patterns of use over the last decade.
- **Parental mediation:** Level of parental engagement, guidance provided at first use, guidance in relation to risks at older ages, rules imposed and limits set.
- **Self-confidence:** Confidence level of children in internet use as perceived by them. Reaction when confronted with an emerging risk, such as where they turn for help and level of knowledge on how to report.
- **Exposure to risk:** Range of risks experienced by children online, type of associated harm, and the way perceived by them. Children's role as *'victim'* or *'observer'* in incidences like online harassment and cyberbullying. Other emerging online risks, such as accepting friends' requests by strangers, exposure to inappropriate content, transmitting very personal or intimate material, etc.

- **Excessive use:** Frequency of online access, type of activity, age of social media engagement and frequency of use, time spend on online gaming and type of activity. Similarities and differences of favorite online activities across children of different educational levels and gender.
- **Involvement of schools:** Involvement of schools in awareness raising, provided education and guidance, readiness to response in emerging risk.

The following step was to identify the interviewing method. Data were collected anonymously with the method of Computer-Assisted Web Interviewing (CAWI), via online questionnaires. Questionnaires were filled online individually by students during their IT course, and under the supervision of their instructor. Specific guidelines were included in the survey design and became known to the instructors that supervised the filling of the CAWI, via online questionnaires. The guidelines were provided to safeguard the proper collection of data, such as privacy during questionnaire filling by students, discrete supervision, no interference by the instructors, etc.

After identifying the method of collecting data, the questions that would cover the above-mentioned thematic areas were formulated. The structure of the questions was analyzed and the wording was depicted carefully as the choice of words and phrases in a question is critical in expressing the meaning and intent of the question to the respondent and ensuring that all respondents interpret the question the same way. Once the survey questions were developed, particular attention was paid to how they were ordered in the questionnaire. Special care was exercised to ensure that the context is similar each time a question is posed, and that the change of thematic areas was communicated appropriately. Last, but not least, the respondents' privacy, ability to remember and time was respected.

Furthermore, the questionnaire has undergone review and acceptance from the [Institute of Educational Policy (IEP)](#), and specifically the Department of Scientific Units. The aim of the IEP is to offer advice and consultation on all matters relating to its fields of study for primary and secondary education with emphasis on curriculum development and design, syllabuses, textbooks, teaching materials and tools, as well as educational research and programs and parliamentary control questions.

The main features in the design of the survey that deserve attention are summarized below:

- The survey involved 400 schools in 2018 and 500 schools the following year, 2019.
- The participating schools were selected from five different cities (urban regions), spread all around the country in 2018 and six different cities in 2019.
- Locations were mostly in mainland Greece, but some locations in the islands and the borderline of the country were also included.
- The survey was approved by the Ministry of Education and Religious Affairs and the involved schools where invited to participate directly by the Ministry following the official avenue. Along with the questionnaire, a set of very specific guidelines was included.
- The questionnaire was filled online, individually and privately by each student, during the IT course under the discrete supervision but not interference of the instructor.
- Demographic features were recorded such as education level (corresponding to specific age group) and gender. The location of the sampling unit (student) and the education level of parents were also recorded.
- The analysis was based on demographic rather than socio-economic features, since the latter did not appear to demonstrate any significant statistical differentiation.

## 2.3 Policy framework and theoretical background of the project

The Greek Safer Internet Center operates with the support of the European Commission, and more specifically the CEF-Telecom programme. According to the CEF-Telecom Safer Internet Workprogram, the main mission of the national Safer Internet Centres is to: *"(i) raise awareness on opportunities and risks that children are facing online, (ii) provide helplines offering support and advice to young people and parents or teachers on problems they encounter online and (iii) operate hotlines taking reports from the public relating to online child sexual abuse material (CSAM) and liaising with law enforcement and ISPs for the swift removal of material found to be illegal."[2]*

Furthermore, *"SICs share tools, resources and good practices and provide services to their stakeholders and users including through the Better Internet for Kids (BIK) platform."* [2] The Greek SIC operates along these three main lines of action, providing an informative and educational portal (www.saferinternet4kids.gr) to raise awareness, a helpline (www.help-line.gr) delivering consulting services and a hotline (www.safeline.gr) offering to the public an online channel to report illegal internet content and activity. The Greek SIC is part of the Insafe network that operates the BIK (www.betterinternet4kids.eu) awareness and resource sharing platform and INHOPE (www.inhope.org) that interconnects all safer internet hotlines operating in Europe and worldwide, liaising with major international stakeholders as well as INTERPOL towards a more effective processing of the reported illegal content and the rapid take down of CSAM material.

As pointed out in the CEF Telecom Workprogramme of the European Commission: *"While the Internet offers many opportunities, it also opens up certain risks to vulnerable users such as children. The Safer Internet DSI (Digital Service Infrastructure) helps to make the internet a safe and trusted environment for children and young users, by providing guidance and awareness raising services and resources at EU level and via interoperable national Safer Internet Centres (SICs)."*[2] The Greek SIC contributes to the European agenda for Internet Safety, acting as the national contact point for the Better Internet for Kids (BIK) Map Policy, which "*…was created to compare and exchange knowledge on policy making and implementation in EU Member States on the themes and recommendations of the European Strategy for a Better Internet for Children (or BIK strategy) first set out by the European Commission in May 2012."*[3] Furthermore, along with representatives from the Greek Ministry of Digital Governance, the Greek SIC represents Greece in the EC Expert Group on Safer Internet for Children, the main advisory body at European level for the online safety of children.

The purpose of the data analysis and the conclusions drawn from the survey is to assist relevant stakeholders, especially national policy makers, to design educational strategies that would allow children, from a very young age, to develop into proper online users and eventually responsible digital citizens. Its purpose is not to ban, depriving children from the wealth of opportunities offered by the online world. On the contrary, the intention of the recommendations is to allow children benefit to the fullest from online activities that enhance their skills and empower their knowledge and creativity.

Education should provide a delicate balance between protective factors and freedoms, between risk and flexibility, offering opportunities for personal development while avoiding overregulation and deprivation. For younger children, more emphasis should be placed to protective factors that would allow them to cultivate their social judgement and develop self-protective skills. For elder children, in their pre-adolescent or adolescent life, protective measures should discretely recede to allow children to develop resilience to risk factors. Adolescence is the time children explore/develop their personality, try to define their place in the world, the time they react to the conventional, sometimes crossing delicate lines, either accidentally, or deliberately due to risky and hazardous behavioral patterns that often jeopardize their safety and put them at risk. The purpose of the educational strategy should not be to deprive them from any of their freedoms, but rather to help them develop self-confidence and resilience, that would allow them to cope with unforeseen situations which could jeopardize their safety and compromise their future.

## 2.4 Ethical aspects

The administration of the questionnaire followed base ethical guidelines, adhering to the national rules and conditions. Furthermore, the questionnaire has undergone review and acceptance from the Institute of Educational Policy (IEP) [4], and specifically the Department of Scientific Units. The aim of the IEP is to offer advice and consultation on all matters relating to its fields of study for primary and secondary education with emphasis on curriculum development and design, syllabuses, textbooks, teaching materials and tools, as well as educational research and programs and parliamentary control questions.

---

[2] https://ec.europa.eu/inea/sites/inea/files/cefpub/c_2020_1078_f1_annex_en_v2_p1_1066015.pdf

[3] https://www.betterinternetforkids.eu/web/portal/policy/bikmap

[4] http://www.iep.edu.gr/en/

In addition, children's anonymity was safeguarded at all times, and the option "I don't know" was included as an answer in all questions. During data collection, special effort was made to provide comfortable conditions for the participants. This included maximizing the anonymity of the participants and discrete supervision - but not interference - of the instructor. Finally, all safety measures to ensure secure communication, protection of the privacy and integrity of the exchanged data were taken. Specifically, the online questionnaire was hosted in a secure environment with HTTPS connection, and the data were stored in an encrypted database.

## 2.5 Limitations

The findings presented in this paper should be interpreted with regard to all limitations that relate to the nature of the data as well as the methodology of the survey:

- Limitations relate to the fact that this is a prevalence and thus observational study that analyzes data from a specific point in time. This kind of studies usually prevent causal inferences.
- Moreover, the study is prone to certain biases, as data are self-reported from children, and trouble with recall should be taken into consideration.

## 3. Internet Use

This section outlines the general attitude of children concerning internet use. Significant parameters in the analysis of the results are the age of children's first online engagement, their means of access, the type of preferred activity by age and gender, the level of information they receive on internet safety and risks, the source of this information, and the degree of parental engagement and supervision.

### 3.1 First engagement and devices used

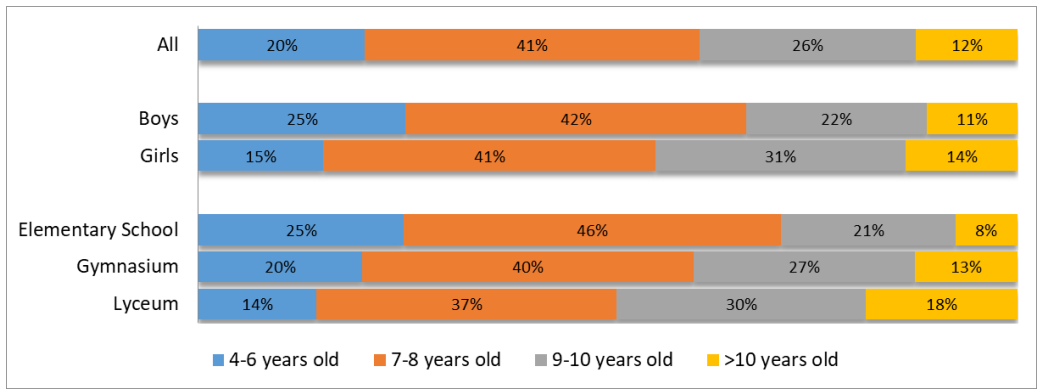

**Fig 3: Question:** When did you start using the internet?

The majority of children (41%) start using the internet at the age of 7-8 years, while 20% of children start at the very young age of 4-6 years. Based on the answers, 25% of boys start using the internet at a very young age (4-6 years old) while 15% of girls start at this age.

From a more thorough analysis of the data, it becomes clear that in recent years children's first contact with the internet occurs at a decreasing age, as indicated by the analysis per education level. That is, 25% of the children currently in the final grades of Elementary school declare that they started using the internet at the age of 4-6 years, while the corresponding rate for children currently in Lyceum is significantly lower (14%) as shown in

Fig 3. At the same time, 8% of children currently at the final grades of Elementary school, state that their first contact with the internet was at above 10 years of age, while this percentage is 18% for students currently in Lyceum.

From the above analysis, it becomes evident that that there is a trend for children to use the internet at a progressively earlier age, starting to access the medium at the very first years of their lives, usually in the preschool period. According to (Coleman & Hagell, 2008) youthful networking is raising many safety concerns, since the young are still developing their social and emotional competencies to manage self-expression, intimacy and relationships.

Based on the literature review study of UKCCIS (Livingstone, Davidson, Bryce, Batool, & Nandi, 2017), the percentage of children using the internet has not changed significantly over the last 5 years, but the amount of time spent online continues to rise steadily.

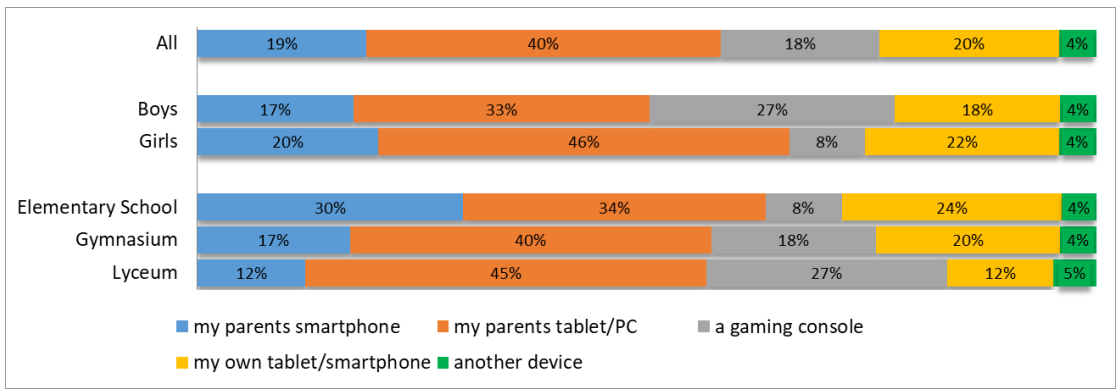

**Fig 4: Question:** What was you first device when started using the internet?

The first device that 40% of children use to surf the web is their parent's tablet or PC, followed by their own mobile phone or tablet at 20%, their parent's mobile phone at 19% and a game console at 18% (Fig 4). These findings are similar to ones reported in (Childwise, 2017) (Ofcom, 2016) (WISEKIDS, 2014) showing that tablets are universally the first device young children use to go online.

A more detailed data analysis reveals that, apart from their parent's tablet/PC, which is unquestionably the predominant answer for children of all educational levels, for younger children their parent's smartphone appears to rank second as their first electronic device, while for older children it appears to be a gaming console. There is an obvious trend here for smartphones to dominate over other electronic devices in recent years, and this same trend emerges in the European *"Study on the impact of marketing through social media, online games and mobile applications on children's behaviour"* (Lupiáñez-Villanueva, et al., 2016).

Moreover, according to (Smahel, et al., 2020) "*In 11 countries (Croatia, Czech Republic, Germany, Estonia, Italy, Lithuania, Norway, Poland, Portugal, Romania and Serbia), over 80% of children aged 9–16 use a smartphone to access the internet at least once a day*".

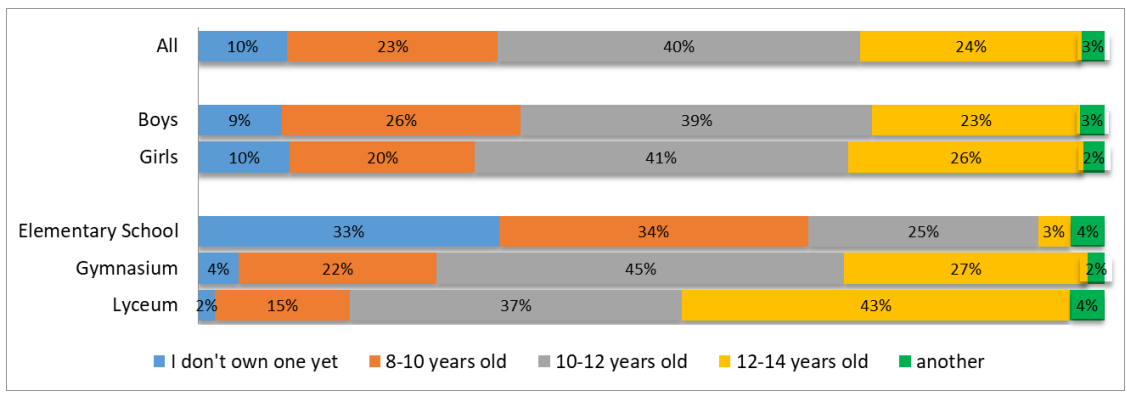

**Fig 5: Question:** At what age did you own a smartphone?

The majority of children (90%) who took the survey own a mobile phone. From them, 40% acquired it at the age of 10-12 years, while 23% acquired it at the age of 8-10 years. Among Gymnasium students, only 4% declare not having a mobile phone (Fig 5), while this same answer was given by 33% of Elementary school students. This means that 67% of children aged 10-12 years old and 96% of children between 13-18 years old already own a smartphone. According to the trends and prediction report of Childwise (Childwise, 2019) 47% of 5-10 year olds now own a smartphone in the UK.

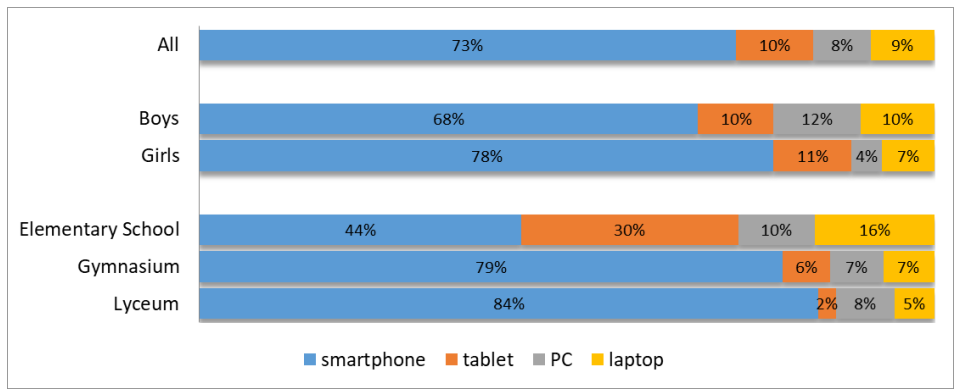

**Fig 6: Question:** What is the main device you use to access the internet?

A mobile phone is the main device used by children to access the internet at 73%. Data show that 84% of students in Lyceum access the web via a mobile phone, 79% in Gymnasium and 44% in Elementary school (Fig 6). From those, 78% of girls and 68% of boys go online mainly through their mobile phones. This 10% difference is mainly because 12% of boys use a PC as the main device, while girls use a PC as the main devices only at 4%. The cause of this difference might be because boys use to play more online games than girls, and it is easier to play on a bigger screen, than a small smartphone screen. Tablets, laptops and PCs are mainly used by Elementary school children at 30%, 16% and 10%, respectively (Fig 6).

According to (Mascheroni & Haddon, 2015) one of the reasons parents buy mobile phones for their children is because the device serves as an "electronic or digital leash" (Ling & Haddon, 2008) that extends parental monitoring outside the domestic environment and keeps track of children's movements. This could also explain the trend witnessed for smartphones to dominate over other electronic devices in recent years rendering internet use for children a very personal experience even from the beginning of their online exploration, as children acquire a mobile phone at a progressively earlier age.

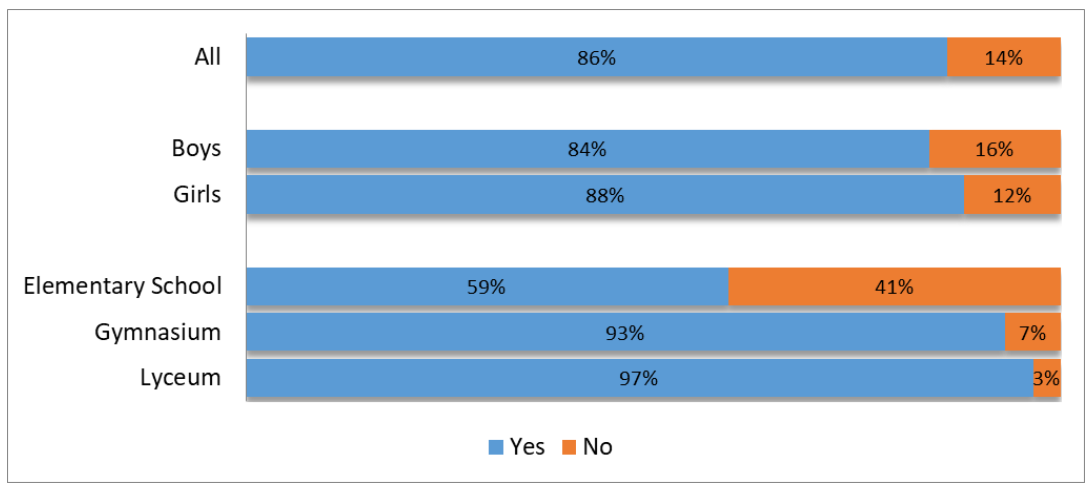

**Fig 7: Question:** Do you use social media (e.g. Instagram, Facebook, Messenger, Snapchat, YouTube, Viber, Skype etc.)?

Children engage with social media from a young age. In the question if they use social media, 86% answer yes (Fig 7). Specifically, Elementary school children (10-12 years old) use social media at 59% although it is prohibited at this young age. Gymnasium and Lyceum students use social media with percentages 93% and 97%, respectively. According to (Ofcom, 2018) report 18% of children in the UK between 9-11 years old have their own profile in social media and for the age group 12-15 years old this percentage rises to 69%.

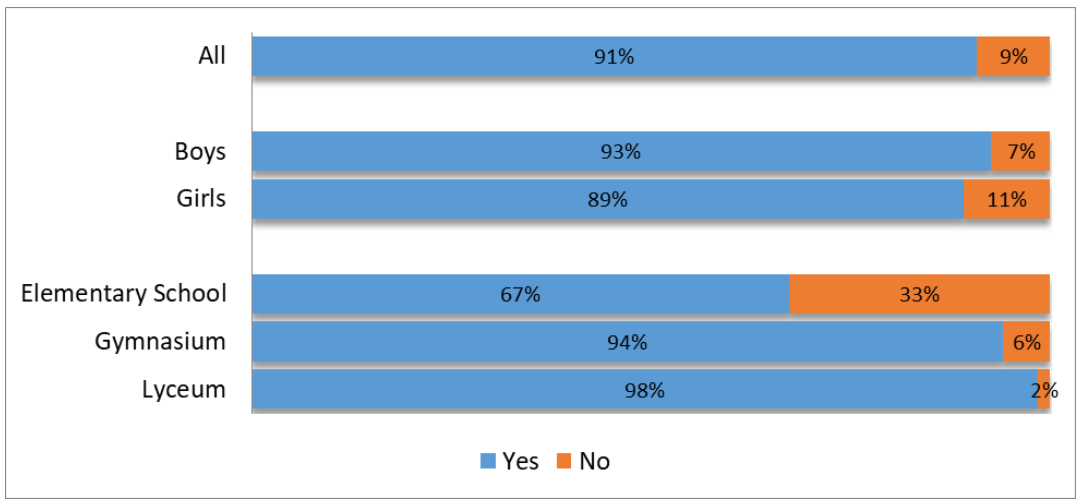

**Fig 8: Question:** Do you have your own profile in the social media you use (not your parents' profile)?

Among the students that answered that they use social media 91% of them answer that they have their own profile and 9% answers that they use another profile (e.g., their parents' profile) (Fig 8). It is also obvious from the responses that children below 13 years of age often create social media profiles (67% of children in Elementary school), although this is prohibited by law. Age falsification is one of the usual tactics children use in order to make their own profiles.

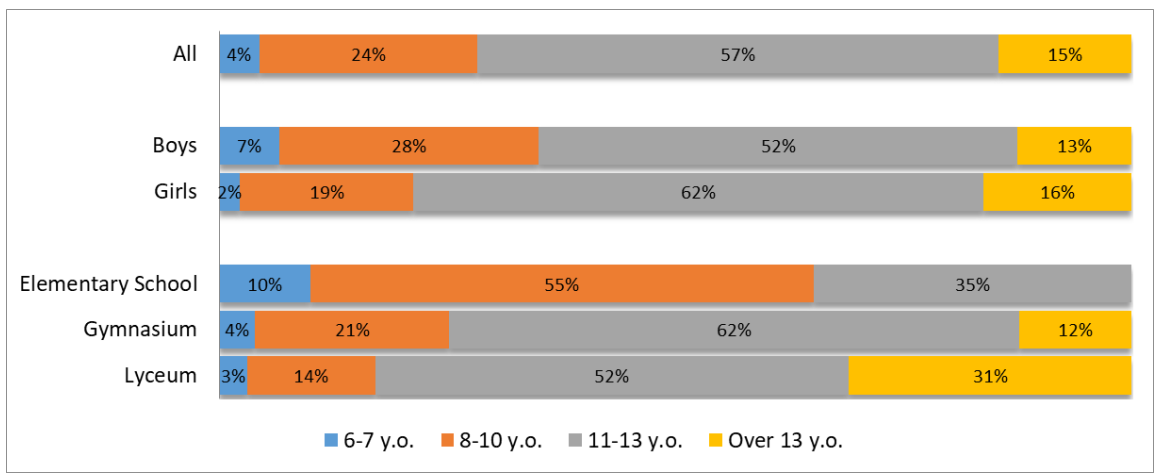

**Fig 9: Question:** When did you start using your social media with you own profile?

In the question, "When did you start using social media with your own profile?" (Fig 9) the vast majority of children from Elementary school answer from the age of 8-10 (55%), but most students currently in Gymnasium and Lyceum answer from the age of 11-13. Furthermore, 31% of students in Lyceum answered over 13 years old. We witness that in recent years children's first contact with social media occurs at an ever-decreasing age. There is also a gender difference between girls and boys. Specifically 35% of boys say that they created a social media profile between 6-10 years old, while for girls this percentage is 21%.

The rates of children 12-14 years old that use social media in other countries of Europe varies between 10% (Finland) and 86% Serbia and Russia (Smahel, et al., 2020). Finally, the percentage of over 14 year olds who use social networking sites daily varies between 21% in Finland and 93% in Countries like Czech Republic and Serbia (Smahel, et al., 2020).

## 3.2 Online behavior and Screentime

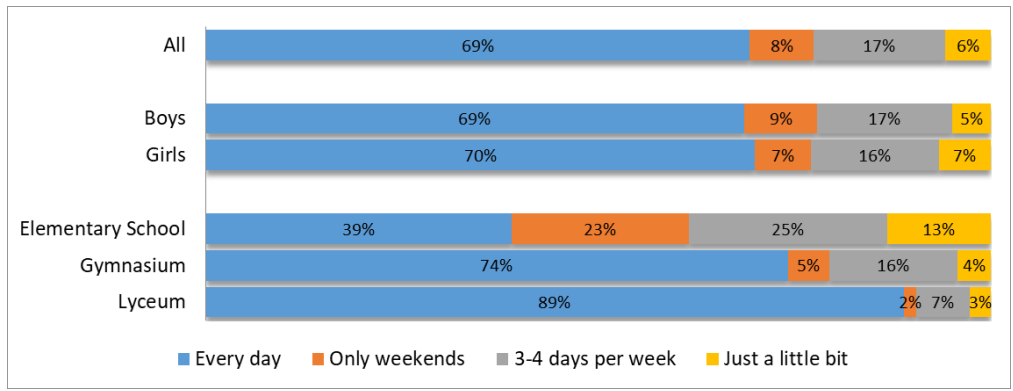

**Fig 10: Question:** How frequently do you use the internet during the week?

Concerning frequency of internet usage, it occurs daily for the grand majority of children. Children engage daily in their favourite online activities that differs as age increases, and between boys and girls. A closer look at the results reveals that 69% of all students access the internet daily, about 17% use it half days of the week, 8% only on weekends and 6% use it just a little. Per education level, 39% of Elementary school students use the internet daily, about 25% use it half days of the week, 23% only on weekends and 13% use it at minimum level. Among Gymnasium students, 74% use the internet on a daily basis, about 16% use it half days of the week, 5% only on weekends and 4% just a little. Finally, among Lyceum students, 89% use the internet daily, 7% about half days of the week, 2% only on weekends and 3% just a little (Fig 10).

It should be noted, that there is direct correlation per education level of the percentage of children that use the internet daily with the percentage of children that use a smartphone as their main device to access the internet. This correlation between everyday use of the internet and smartphones could be explained from the fact that smartphones offer seamless connectivity anywhere and anytime and have already become children's ultimate personal devices.

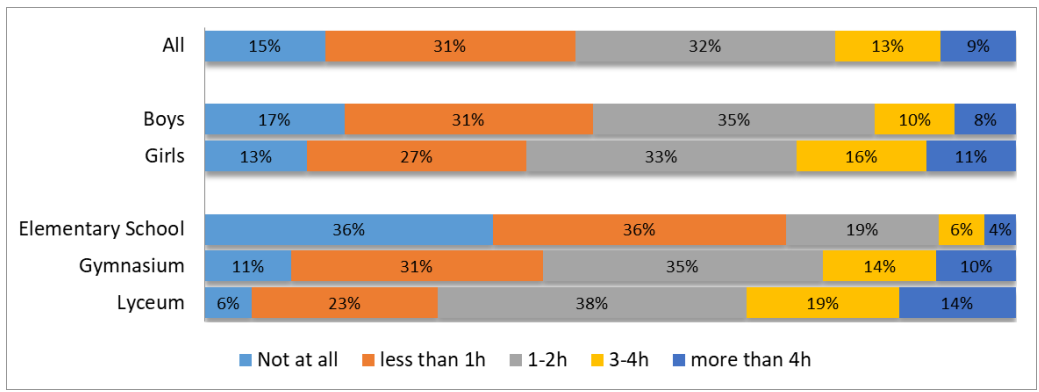

**Fig 11: Question:** How many hours do you spend on the internet daily?

15% of all students say that they don't spend time on the internet daily, about 31% use it less than 1 hour, 32% use it 1-2 hours, 13% use it for 3-4 hours daily while 9% more than 4 hours daily. The majority of Elementary school student replied that either they do not use it daily (36%) or they use it less than 1 hour (36%). On the contrary, the majority of Gymnasium and Lyceum student said that they use it 1-2 hours per day (35% and 38% respectively). It is also interesting that 24% of Gymnasium students and 33% of Lyceum students say that they are online for more than 3 hours per day (Fig 11).

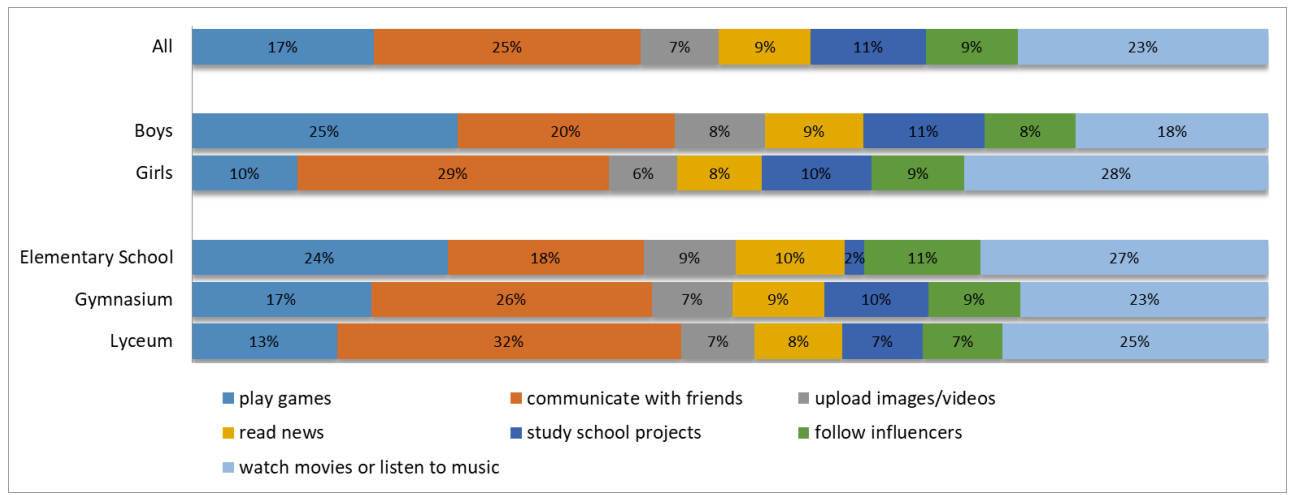

**Fig 12: Question:** What do you more frequently do when online?

In the question, "What do you more frequently do when online?" (Fig 12) children of all ages respond they communicate with their friends (25%), watch movies or listen to music (23%), play games (17%). Girls mostly communicate with their friends (29%), watch movies or listen to music (28%) while boys mostly play games (25%). As expected, there is a clear gender bias related to the online activities of children. Playing games dominates boys' online activity (25%) as opposed to communicating with friends (20%) and watching movies/listening to musing (18%). This conclusion is in accordance with the study of EU Kids Online from 11 countries of Europe (Smahel, et al., 2020). More specifically, in their survey it is noted that around twice as many boys as girls play games online daily, a conclusion also derived from our study.

Girls' online time is equally spend between communication with friends (29%) and watching movies/listening to music (28%) rather than game playing (10%). The remaining activities polled (upload images/videos, read news, study schoolwork, follow influencers) score equally for both genders.

Watching movies and playing games comes first in the preference of Elementary school children (27% and 24% respectively), while the favourite online activity of Gymnasium and Lyceum students is communicating with their friends (26% and 32% respectively) (Fig 12). Children from all educational levels enjoy watching movies/listening to music on the internet. As children grow older, communication becomes their predominant online activity while playing games declines. This has roots to the fact that media offer children orientation and the potential for identification (Livingstone S. a., 2001). Furthermore, media can have a supporting function in young people's socialization and further the development of social understanding (Livingstone, d'Haenens, & Hasebrink, Childhood in Europe:, 2001) (Paus-Hasebrink, Kulterer, & Sinner, 2019).

Similar trends are reported from Childwise (Childwise, 2017). More specifically, Childwise's Monitor Report reveals that children aged 7-16 use the internet to watch video clips (59%), listen to music (56%), play games (54%), complete homework (47%), interact with family and friends (47%), social networking (40%), look up information (38%), and upload videos, photos and music (27%).

## 3.3 Parental mediation

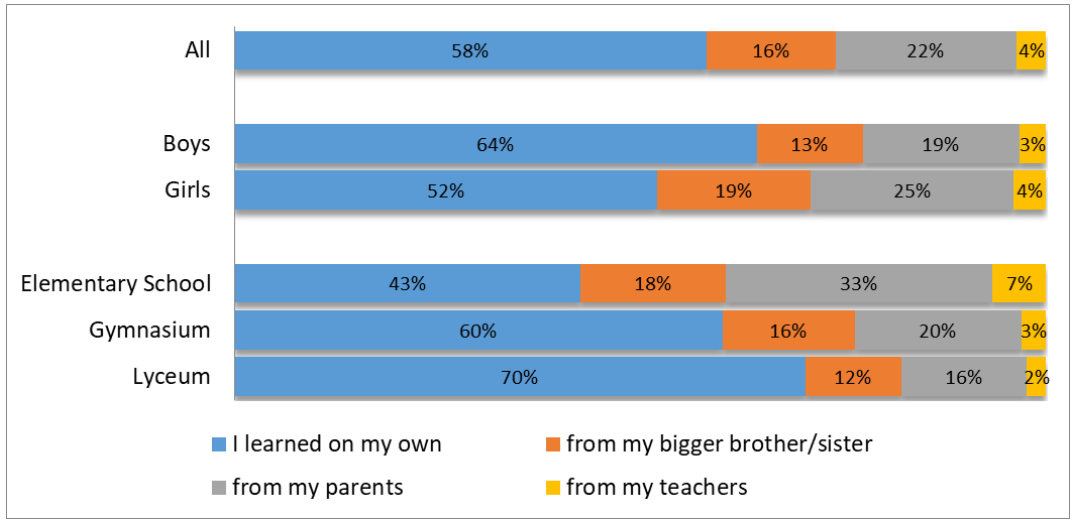

**Fig 13:** Did you get any guidance in the process of learning how to use the internet?

Our results reveal, that the majority of children (58%) declare they did not get any guidance in the process of learning how to surf online. A percentage of 22% state they had the help of their parents, 16% from an older sibling and only 4% got help from school or an educator (Fig 13). From a more thorough analysis of the results, it becomes evident that, in the course of time, children continue largely, to start using the internet on their own. Primarily parents/caregivers and secondarily older siblings take over the task of guiding children in this endeavor. According to the answers, only 7% of Elementary School, 3% of Gymnasium and 2% of Lyceum students had the assistance of an educator when they first started using the internet (Fig 13).

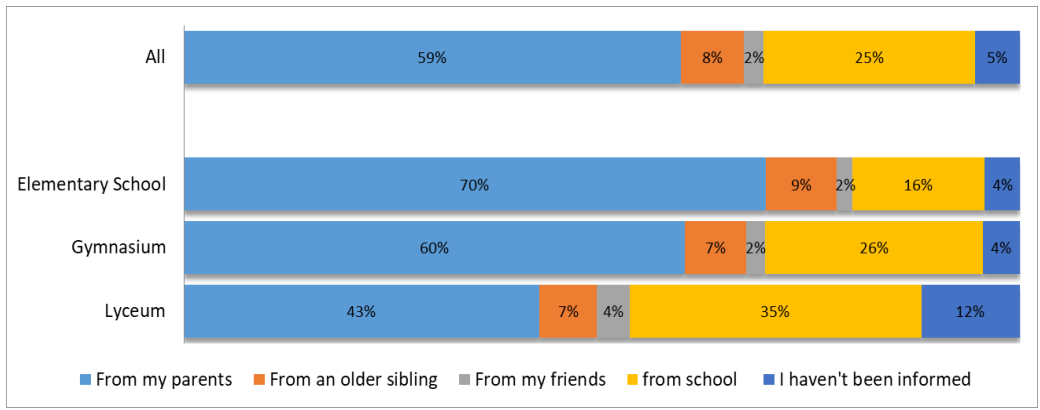

**Fig 14: Question:** From whom/where have you been informed about safer internet use?

On the question "From whom/where have you been informed about safe internet use?" (Fig 14) students mainly answer from their parents (59%) and at a smaller percentage from their school (25%). 5% of the children say that they have not been informed at all, an 8% declare to be informed by their elder sibling, and a 2% by their friends.

Curiously enough, as time advances children get increasing information from parents, an encouraging fact, since this demonstrates increased awareness and involvement of parents. The information children get from school seems to be at a lower percentage for elementary school children (16%), more in gymnasium (26%) and even more in lyceum (36%), which can be interpreted by the fact that internet education comes at later stages, giving emphasis to children in their adolescent rather than early ages.

Unfortunately, internet safety has not been essentially incorporated into the school curriculum, but occurs spontaneously, and not in an organized manner. It occurs from informative sessions of the Greek Safer Internet Center, or online safety material exploited by the Gymnasium and Lyceum professors. These observations, in combination with the earlier occupation of children with the online world, dictate the need for a more organized treatment of the topic by the educational systems, starting education in schools in an organized way, incorporating material into school curricula, from as early as kinder garden and even earlier, at the preschool period.

Ofcom's survey (Ofcom, 2018) also shows the perspective of the parents. According to this report, nearly all parents of 3-4s and 5-15s (97%) mediate their child's use of the internet in some way, either through technical tools, supervision, rules or talking to their child about staying safe online. Only a minority of parents (between 21% and 32%) whose child has a profile in a social media platform, knows and can state correctly the minimum age for Facebook, Instagram and Snapchat.

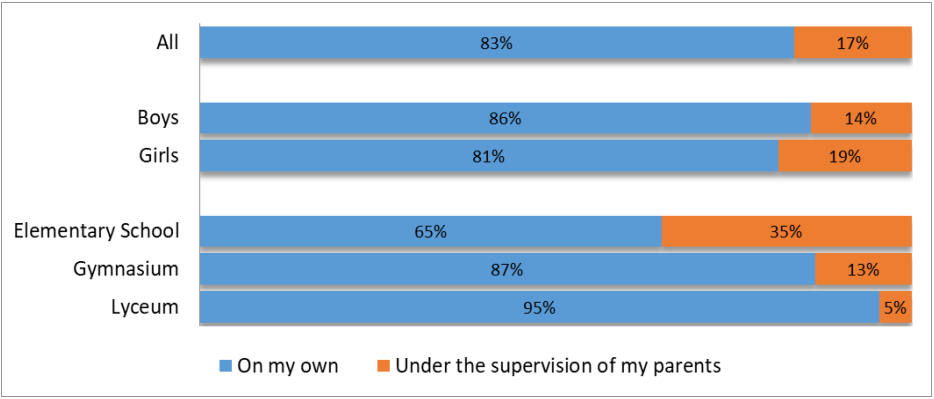

**Fig 15: Question:** Are you surfing the internet on your own or under the supervision of your parents?

From the children, 83% (50% boys - 50% girls) state they go online without any supervision, and a deeper analysis of the results reveals that even for children of Elementary school there is no sufficient parental supervision when online. More specifically, from the respondents, 65% of Elementary school, 87% of Gymnasium and 95% of Lyceum students, do not have any supervision whatsoever regarding internet use (Fig 15).

Complementary to that, Ofcom's report (Ofcom, 2018) shows that parents often feel unable to follow their kids' online activities. For example it is stated *"Sarah's mum generally checked up on what Sarah's younger siblings did online, but she felt that she did not need to check what Sarah (aged 15) did, as she was old enough to make her own decisions. In any case, she felt there was no way of stopping Sarah having social media accounts. She explained that she just has to trust her"*.

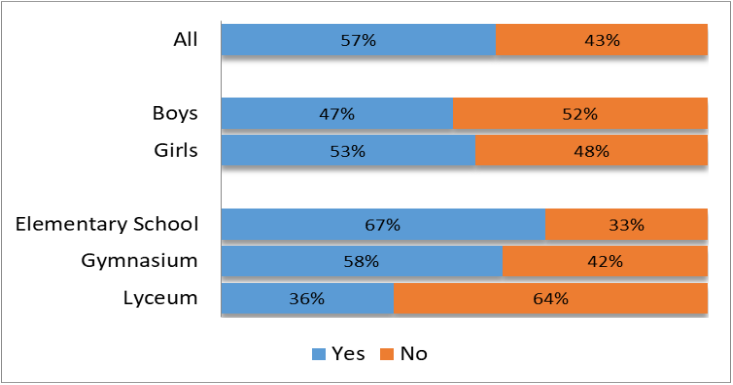

**Fig 16: Question:** Do your parents impose you any limits regarding the use of the internet?

As a result of all children's answers, almost half of parents (43%) do not set limits for their children (48% for girls - 52% for boys) regarding internet use. A deeper analysis by educational level reveals that, for 33% of Elementary school, 42% of Gymnasium and 64% of Lyceum students, parents do not impose any rules regarding internet use (Fig 16). Lack of parental mediation and supervision, as well as very little to not at all regulation, signifies the need for enhanced parental awareness.

One of the main outcomes of the survey was that most parents seem to ignore the importance of setting limits and rules concerning internet use to their kids from very young age, and they lack the required vigilance on this topic. Indeed according to Ofcom's survey (Ofcom, 2018), were parents were asked, despite rising concerns, parents are, in some areas, less likely to moderate their child's activities. It is also reported that some parents were resigned to feeling they had limited control over their children's online activities, and this was especially applicable for older children.

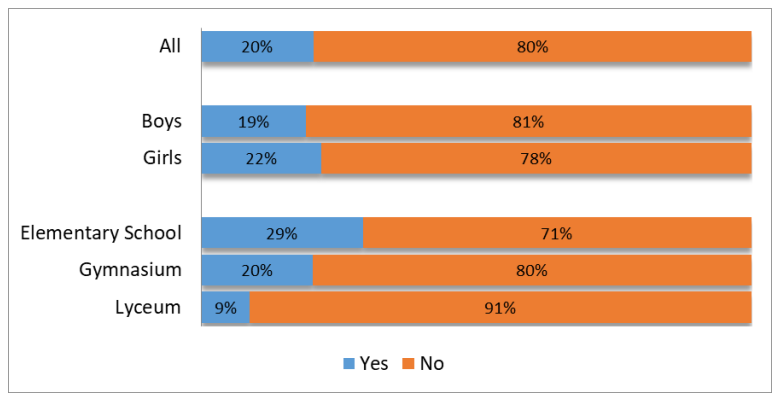

**Fig 17: Question:** Are your parents using any parental control system?

According to the answers of the children, 80% declare that their parents do not use any parental control system. For children of Elementary school, this percentage is 71% and only 29% of the children say that their parents use parental control systems. As children grow, we observe that the percentages of parents using parental control systems decline even more. Specifically only 20% of children of Gymnasium and 9% of children in Lyceum say that their parents use parental control systems.

In previous questions of this section, we explored active mediation of parents when it comes to internet safety (e.g. when a parent discusses and explains online activities and content with the child), while in this last question we have explored the monitoring means of parents and caregivers (e.g. when a parent checks on the child's online activities) (Talves & Kalmus, 2015). Our study shows that parents who engage with active mediation, usually also monitor their children's online activity.

## 3.4 Confidence level of children

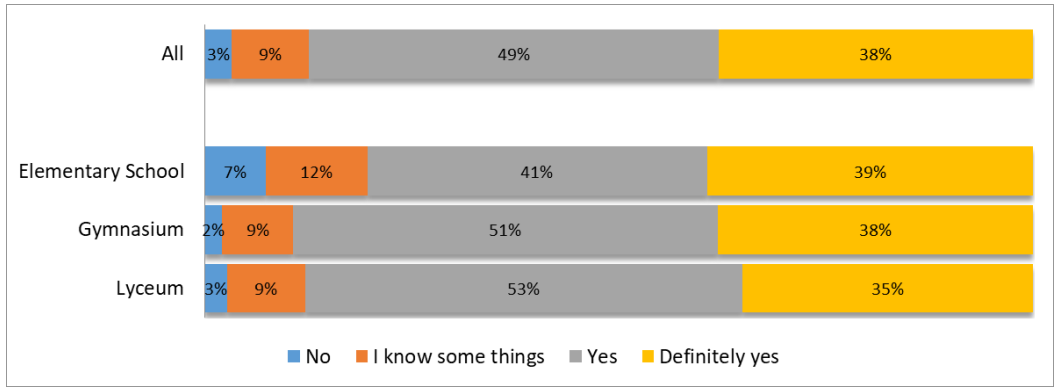

**Fig 18: Question:** Do you believe you know how to use the internet safely?

The grand majority of children declare that they feel confident, to a large extend, that they are responsible internet users and they know how to report something inappropriate that is found online.

Despite the fact that most children declare that did not get any guidance when first started to use the internet (Fig 13), and that they go online without supervision (Fig 15), they feel confident they know how to use the internet safely and only 12% said that they are not educated (answered "No" and "I know some things") (Fig 18). This percentage is 19% for Elementary school, 11% for Gymnasium and 12% for Lyceum students. On the other hand, children that answered "Yes" or "Definitely Yes" are 80% in Elementary school, 89% in Gymnasium and 88% in Lyceum.

One would anticipate that as children grow up, the more confident they would feel about their knowledge on how to keep safe online. However, we notice that this is not the case in Lyceum. In combination with the question "From whom have you been informed about safer internet use?" (Fig 14) we derive the conclusion that children in Greece start to get some kind of formal education on internet safety in Lyceum, and that is the also time they start to question their knowledge about internet safety.

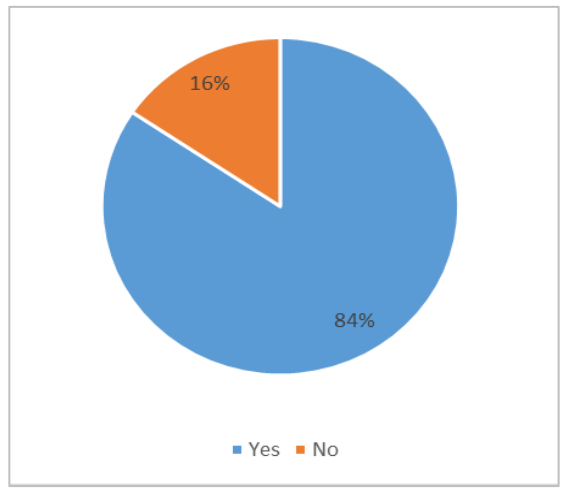

**Fig 19: Question:** Do you know how to report something/someone that upset you online?

In Fig 19, we see that 16% of children do not know how to report something that upset or disturbed them online, while 84% of them stated that they know how to do so. In the 2016 Ofcom survey (Ofcom, 2016), the percentage of children declaring that they know how to report something upsetting they found online is 32%, but this percentage increased to 70% in the 2018 Ofcom survey (Ofcom, 2018) for 12-15 year olds. Our findings are in line with the latest, 2018 Ofcom survey (and there is a big gap with the findings reported in Ofcom 2016) which may be explained from the fact that there is an obvious trend in recent years to put effort into awareness raising and online reporting methods.

Although children appear to be confident about their knowledge on internet safety, it is not certain that this reflects reality given the risks they engage in. *Adolescent is the period for children to explore, define their identity and develop resilience. Guidance from an early stage should be a prerequisite in order for the beneficial aspects of online activity to prevail over potential unsafe engagements. Awareness and vigilance should be strengthen from the very early stages of children's digital lives, when exploration still takes place in a more controlled environment.*

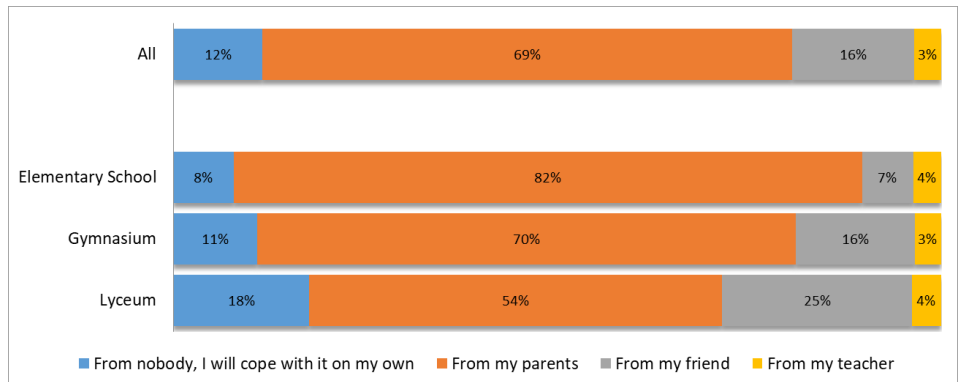

**Fig 20: Question:** If something serious happens to you online, to whom would you turn for help?

In the question "If something serious happens to you online, to whom would you turn for help?" 69% responded to my parents, 16% to a friend, only 3% to a teacher, and there is a 12% stating: "No one, I'll deal with it on my own". A very high percentage of Elementary school (82%) and Gymnasium students (70%) said that they would turn to their parents for help. However, this is not the case for older children as the percentages drop to 54% for Lyceum students' who stated they would ask for help from their parents, 25% from a friend and 18% would try to deal with it alone (Fig 20). The results of the survey highlight that, on average, only a 3% of students from all three educational levels would turn to a teacher for help. This is a fact calling for attention, as far as education in Greece is concerned, as apparently, not enough effort is devoted to internet safety awareness, and there is no structure provided by schools where children could turn for guidance and advice.

Furthermore, children largely declare that they would seek help from parents if they experience problems online, which is promising and shows that the family structure in Greece is still very strong. At increasing percentage, as we go from Elementary school to Lyceum, children declare that they will try to cope with the situation on their own, which is natural since older children develop resilience. This percentage however is still low in all educational levels.

An alarming observation is drown from the fact that school appears to be non-existent in coping with emergency/anxiety online occurrences, explained by the fact that schools in Greece lack, in general, the required structure to deal with emergency situations. School psychologists were appoint for the first time, experimentally, during academic year 2018-19 as pilots in selected school units. Emergencies are usually left to educators to deal with in a spontaneous manner, which apart from the fact that they do not possess the proper training; their main responsibility is the cognitive part of the education and not the psychological aspect. Schools lack the mechanisms both to identify alarming situations but also to deal with them.

## 3.5 Self-assessment and self-image

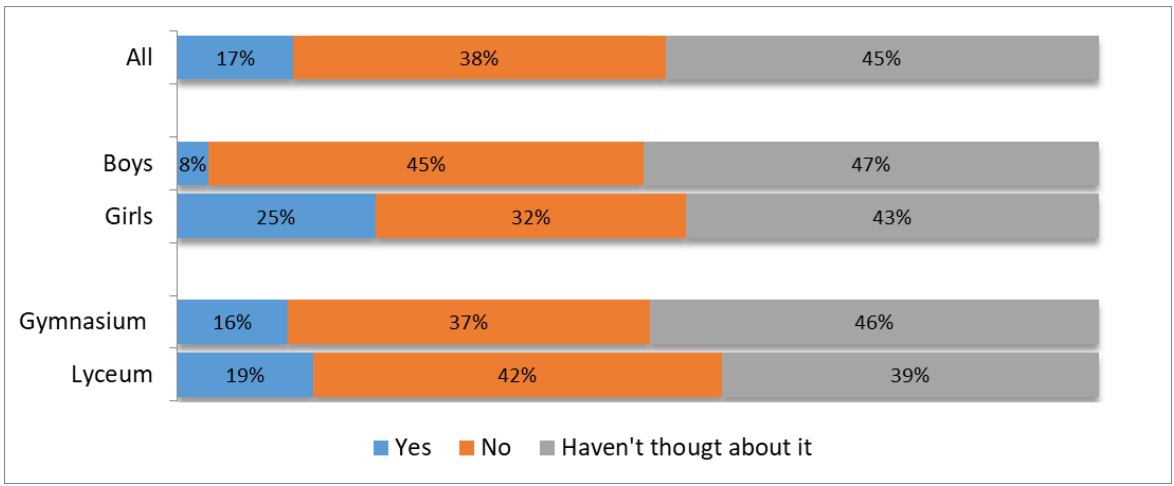

**Fig 21 Question:** Do you think that you are unattractive when you compare yourself with images of other online?

A small percentage of children 12-17 years old believe that they are "unattractive" compared to pictures of others that they see online, but the majority of children either hasn't thought about it, or does not believe this is true. There is an obvious gender difference when it comes to body image and self-confidence online. Namely, one out of four girls do compare themselves with images of people they see online and feel less about themselves (Fig 21), while this is the case for only 8% of boys.

There does not seem to be any difference between been affected from images of friends or images of influencers they follow (Fig 22). Most of them say that their self-confidence is not affected (68% of Gymnasium and 67% of Lyceum students). There is a relatively big percent stating that they are affected just a little (27% from friends and 26% from influencers). We can again observe a gender difference as girls more than boys tend to be affected (by friends 31% girls vs. 22% boys, by influencers 31% girls vs. 21% boys).

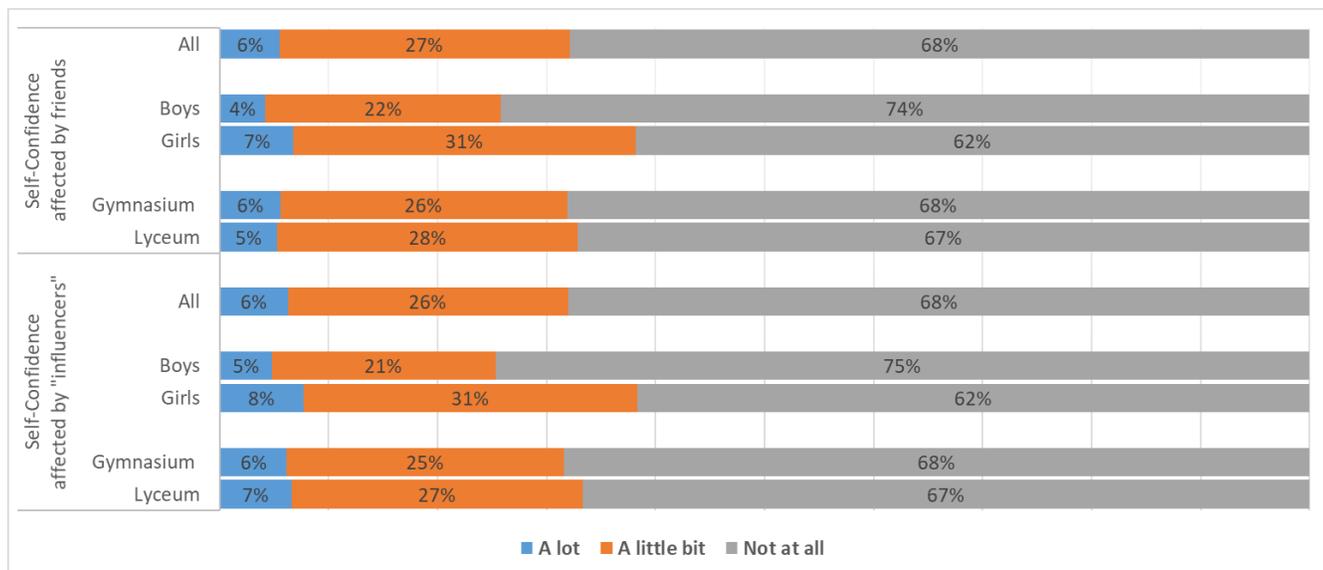

**Fig 22 Question 1:** How much does the photos/material uploaded by your friends affect your self-confidence?
**Question 2:** How much does the photos/material uploaded by influencers you follow affect your self-confidence?

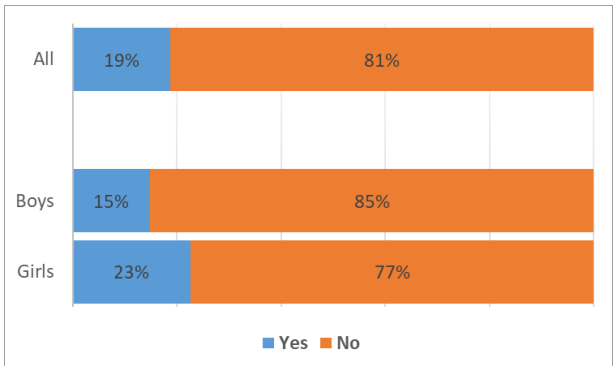

**Fig 23 Question:** Have you ever deleted an image/video that did not get many likes?

Furthermore, adolescence (12-18 years old) seem to be interested on the number of likes they would receive when they upload something in social media platforms. 24% said that they are worried about the number of likes they will get. As expected, here there is also a gender difference, with 29% of girls saying they are worried about peers' approval, while for boys the respective percentage is 10% less (Fig 24).

23% of adolescent girls (12-18 years old) say that they have deleted an image/video that did not get many likes, while for boys the respective percentage is 15% (Fig 23).

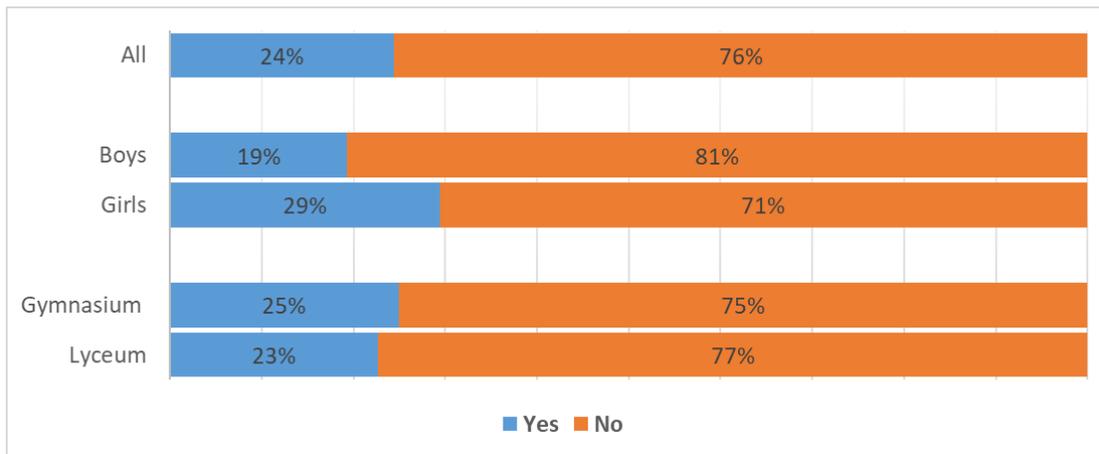

**Fig 24 Question:** When you upload a photo of yourself online, are you worried about the number of likes it will get?

### 3.6 Digital literacy

While children come to understand the digital environment better with age and experience, it is by no means clear that a critical understanding of the digital environment results in cautious behavior regarding personal data protection. Indeed, data reveal how uneven children's digital literacy is. Having said that, the results of our survey show that 44% of children (54% boys - 46% girls) do not think about the impact a photo or a video they share could have on their online reputation (Fig 25).

About half of them (22%) believe they can permanently delete material published online at any time (Fig 25). Out of Elementary school children, 15% have the misconception they can delete material published online, 25% of Gymnasium students and 20% of Lyceum students. At the same time, in the question "Do you know how to use the internet safely?" 88% answered that they know and only 12% that they don't know (Fig 18).

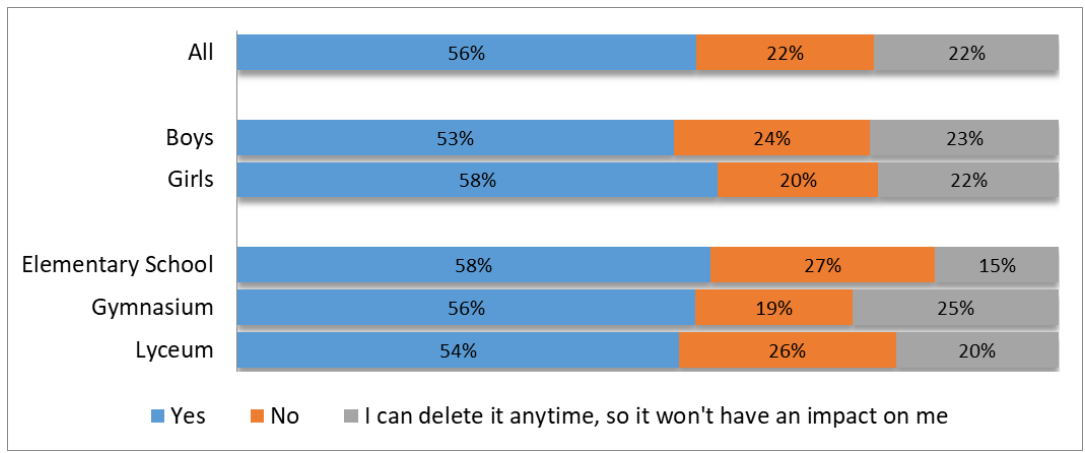

**Fig 25: Question:** Do you think about the impact a photo or a video could have on your online reputation before posting it?

Furthermore, children report that their main source of information about how to keep safe online are predominately their parents (59%), then the school (25%), followed by an older sibling (8%), and 2% a friend, while 5% reports that they have not been informed. Unfortunately, as stated above, in Greece, there is no dedicated lesson or curriculum for internet safety, and thus children do not get the adequate and appropriate information from school. Teachers mainly make use of the resources provided by the Greek Safer Internet Center, however, public education should also incorporate digital literacy curriculums is schools.

Focusing on the use of social media, Livingstone (Livingstone S. , Developing social media literacy: How children learn to interpret risky opportunities on social network sites, 2014) shows that children's digital literacy changes qualitatively with age:

- at around 9-10, children are concerned with what's real or not, although they may not discriminate real from fake clearly;
- at 11-13 they are more concerned with what is fun or even transgressive, irrespective of whether it is trustworthy;
- by 14-16 their increasing maturity leads teenagers to refocus on what is more valuable for them or more generally.

From the Ofcom's surveys (Ofcom, 2016), (Ofcom, 2018) it is also obvious that children's digital literacy increases steadily from age 8 to young adulthood. For example, it becomes gradually less likely to think that all information on news media sites is true. This is something also shown in our survey. Namely in the question "Do you think that you can recognize a true story from fake news?" more than half (58%) of children believe that they can distinguish real from fake news (Fig 26). A higher percentage of Lyceum students answered "yes", compared to Gymnasium and Elementary school students. More specifically:

- 65% of Lyceum students, 59% of Gymnasium students, and 51% of Elementary school students answered that they can recognize the difference;
- 27% of Lyceum students, 31% of Gymnasium students, and 29% of Elementary school students answer that they can recognize it only if it is very obvious;
- while 8% of Lyceum students, 10% of Gymnasium students, and 20% of Elementary school students answer that they are not able to recognize the difference.

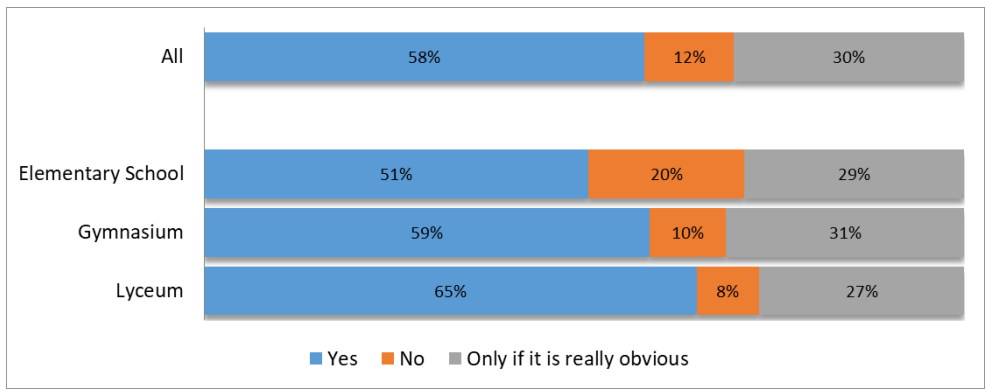

**Fig 26: Question:** Can you recognize fake news online?

Moreover, in the question "What is the main source you get the news from?" children and young people answered that they mainly get their information from publications on social media (36%) and from friends' posts (15%). Lyceum and Gymnasium students answered mainly from social media posts (48% and 39%) and secondly from news sites (30% and 26%), while Elementary school students answered mainly from other sources (39%) and secondly from news sites (25%) (Fig 27).

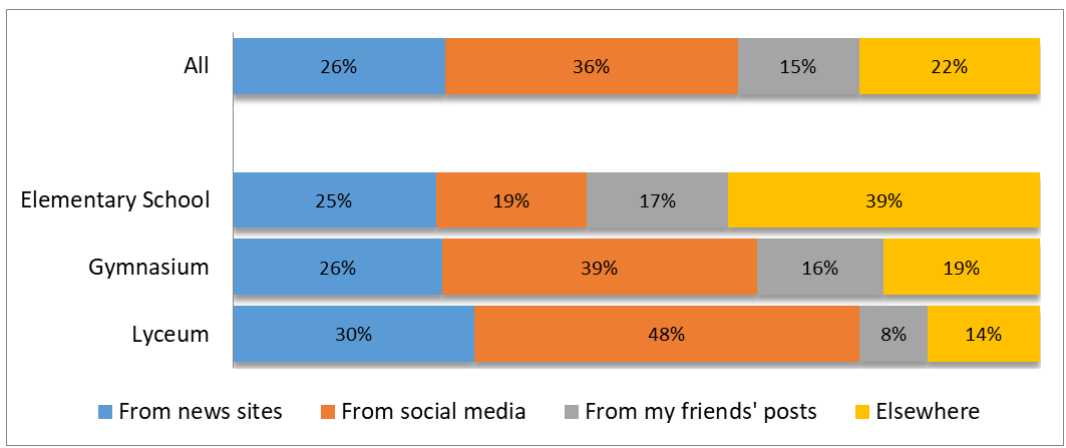

**Fig 27: Question:** From which news source do you get your information?

## 3.7 Social Media use and privacy

The relation of children with social media and the adopted use policies were included in this section part of the survey. Social media is a term for the online platforms that children and adults use to connect with others, share media content, and form social networks. Using social media means uploading and sharing content. Although social media were designed for and primarily are used by adults, children and young people have taken up social networking with alacrity, and this is reshaping youthful practices of communication, identity, and relationship management (Livingstone, Ólafsson, & Staksrud, Risky Social Networking Practices Among ''Underage'' Users: Lessons for Evidence-Based Policy, 2013).

A systematic narrative review investigating the effects of online technologies and social media on adolescent mental well-being (Best, Manktelow, & Taylor, 2014) reported that the benefits of using online technologies were increased self-esteem, perceived social support, increased social capital, safe identity experimentation and increased opportunity for self-disclosure. On the other hand, harmful effects were increased exposure to harm, social isolation, depression and cyber-bullying.

The age of consent for social media and the conditions under which this engagement begins are the two basic parameters examined. The vast majority of the children of our survey (86%) has a profile on a social network, of which 66% created a profile before the age of 13 years old (Fig 30). In the 2018 Ofcom survey (Ofcom, 2018) it is reported that only 18% of 8-11 year olds and 69% of children aged 12-15 years old have their own social media profile. Back in 2013, according to Livigstone's et.al. (Livingstone, Ólafsson, & Staksrud, Risky Social Networking Practices Among ''Underage'' Users: Lessons for Evidence-Based Policy, 2013), 38% of 9-12 year olds who use the internet in the 25 European countries have their own social media profile. NSPCC findings (Lilley & Ball, 2013) suggest that 59% of UK 11-12 year olds have social media accounts.

The question "Did you open an account on your own or with the approval of your parents?" shows that 34% of children with active social media profiles created them on their own without the consent of their parents (Fig 28).

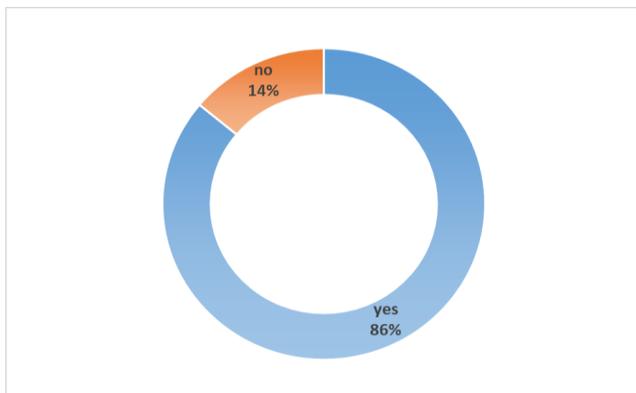
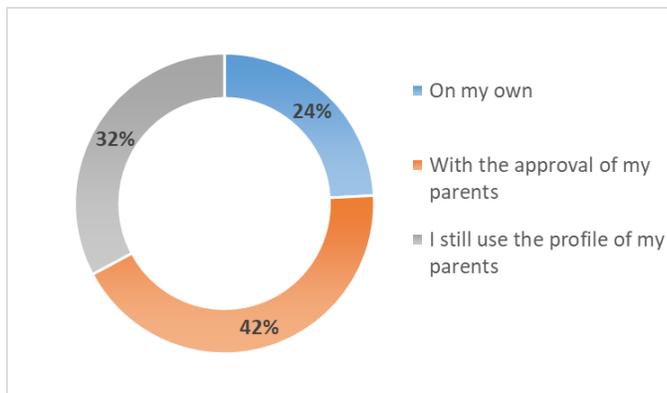

**Fig 28: Question:** Do you have account in a social media

**Fig 29: Question:** Did you open your account on your own or with the approval of your parents?

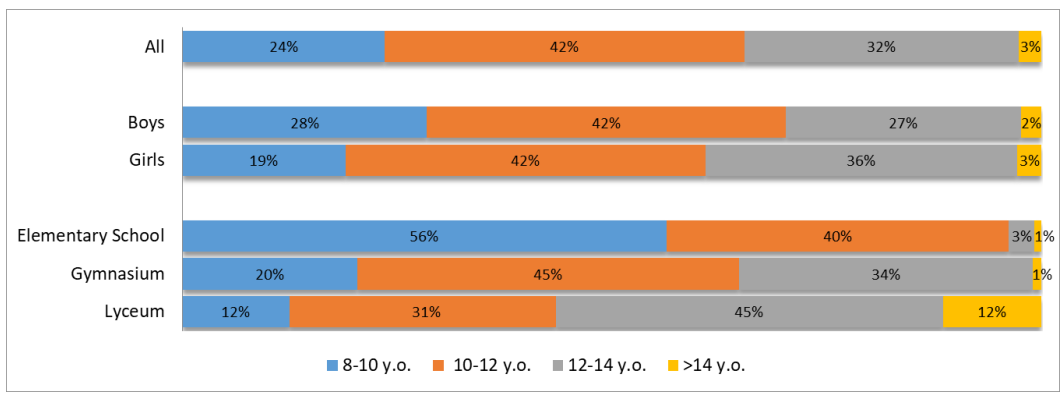

**Fig 30: Question:** At what age did you start using social media?

It is clear from the responses that children who answered that they have an account, use social media at a progressively decreasing and non-permissible age. More specifically, 96% of the children in Elementary school, 65% of Gymnasium and 43% of Lyceum students declare to have entered the social networks at a non-permissible age (Fig 30). Of the children who responded that they started using social media at the age of 8-10 and 10-12, 48% are boys and 52% girls, thus there is no apparent gender difference.

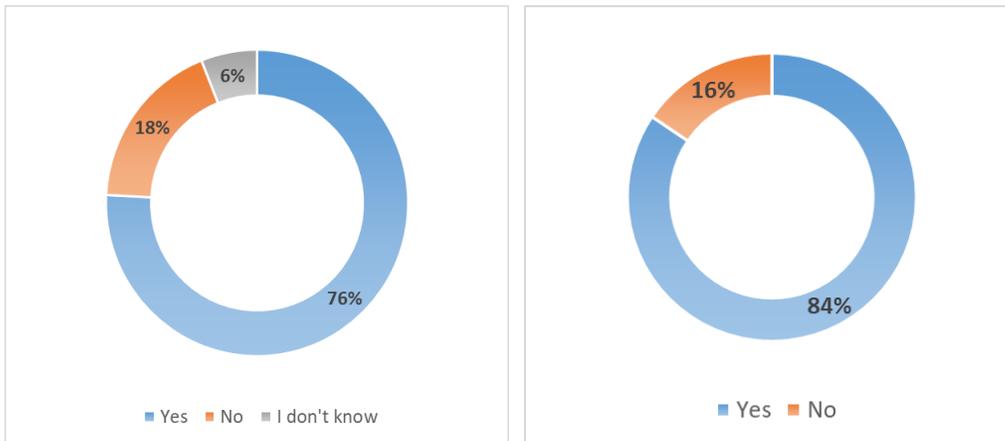

**Fig 31: Q:** Do you have your profile private on social media? **Fig 32: Q:** Do you know how to report something that upset you?

From the children that make use of social media, 18% have not set their profile to private, 76% has set it to private, and 6% does not know what a private profile is (Fig 31). 84% of the children said that they know how to report somebody/something that upset them online and 16% stated that they do not know how to do so (Fig 31). In the 2016 Ofcom report (Ofcom, 2016), 36% of children report that they know how to change the settings so that fewer people can view their social media profile, however only 18% has actually done it.

Instagram is the social network used more frequently for Gymnasium and Lyceum Students (63% and 71%), while for Elementary school students it is YouTube (35%). The second most used social media for Gymnasium and Lyceum Students is YouTube (22% and 20%), while the second most used for Elementary students is Viber (23%), followed by TIK TOK (21%). For the general population, after Instagram (57%) and YouTube (24%), follows Viber (8%), then TIK TOK (7%), and then Snapchat (1%) and Facebook (1%) (Fig 33). In contrary with the results of our survey, the Ofcom 2018 survey (Ofcom, 2018) shows that the most popular social media in the UK is Facebook (31%), and then Snapchat (31%) and Instagram (23%), for the 12-15 year olds.

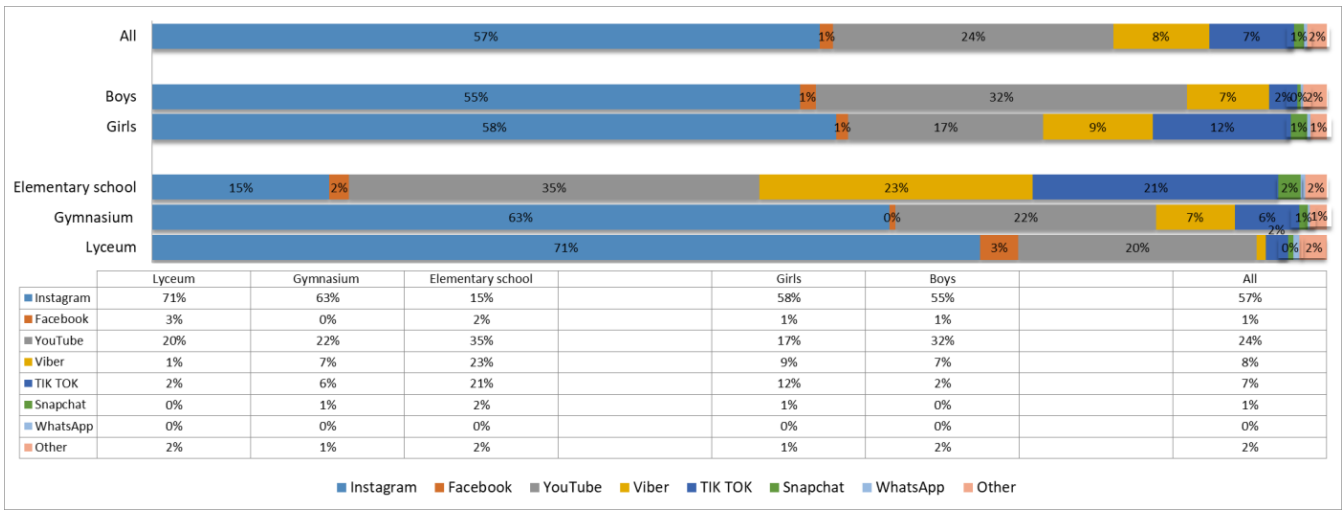

|  | Lyceum | Gymnasium | Elementary school |  | Girls | Boys |  | All |
|---|---|---|---|---|---|---|---|---|
| Instagram | 71% | 63% | 15% |  | 58% | 55% |  | 57% |
| Facebook | 3% | 0% | 2% |  | 1% | 1% |  | 1% |
| YouTube | 20% | 22% | 35% |  | 17% | 32% |  | 24% |
| Viber | 1% | 7% | 23% |  | 9% | 7% |  | 8% |
| TIK TOK | 2% | 6% | 21% |  | 12% | 2% |  | 7% |
| Snapchat | 0% | 1% | 2% |  | 1% | 0% |  | 1% |
| WhatsApp | 0% | 0% | 0% |  | 0% | 0% |  | 0% |
| Other | 2% | 1% | 2% |  | 1% | 2% |  | 2% |

**Fig 33: Question:** Which social media platform do you use more frequently?

Regarding Video Sharing Platforms (VSPs) like YouTube/Vimeo/etc., children answer that they do watch videos on them 96%. Watching videos on VSP is so widespread, that all educational levels engage with it. Therefore, 96% of Elementary school children, 96% of Gymnasium children and 98% of Lyceum students watch videos on VSPs (Fig 34). Very high percentages are also reported in the Ofcom 2018 survey (Ofcom, 2018), namely 77% of 8-11 year olds and 89% of 12-15 year olds using YouTube.

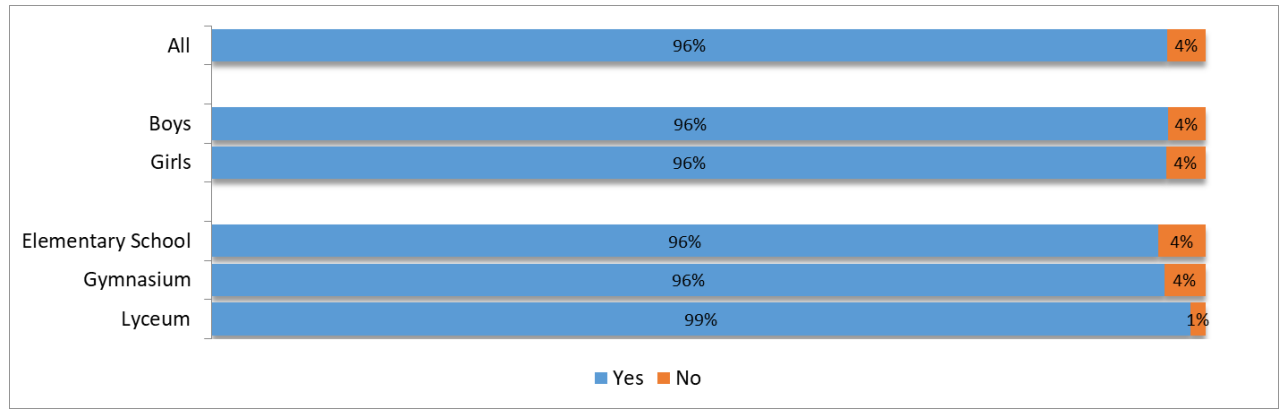

**Fig 34: Question:** Do you watch videos on video sharing platforms like YouTube?

Regarding the question what they like to do on VSPs (Fig 35) children answer that they like to listen to music (60%), follow YouTubers (59%), watch videos/movies series (35%), watch tutorials (18%), and upload their own videos (6%). Boys seem to enjoy more to follow YouTubers (69%) followed by listening to music (50%), while girls say that they like more to listen to music (69%) followed by following YouTubers (50%). It seems that boys create and upload their own videos more than girls (10% vs. 2%) so boys tend to be a more active audience on VSPs compared to girls. On the other hand, girls more than boys prefer to watch videos/movies/series (43% vs. 27%).

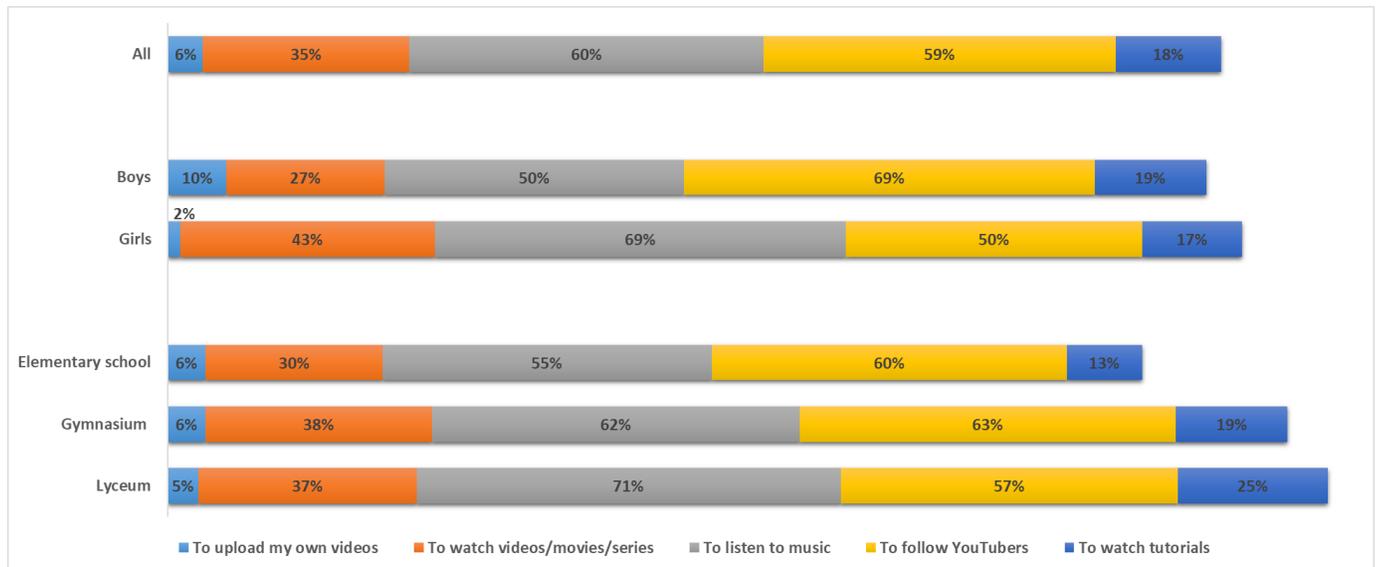

**Fig 35: Question:** What do you like to do when you are surfing on VSP like YouTube?

## 4. Online Risks

A large part of our survey is devoted to online risks and the extent to which children can identify them properly and protect themselves. Cyberbullying, grooming, exchange of personal photos (sexting), and disclosure of personal data, online reputation and misinformation are the main risks explored. Through the answers of the students, important conclusions are drawn that can assist in raising awareness on children, parents and educators. It should be explicitly mentioned that online activities are inherently neither beneficial nor harmful; rather, outcomes depend on the nature, the frequency, the context of use (privacy settings, digital literacy level, safety skills) of the internet (Livingstone, Haddon, & Görzig, Children, risk and safety on the internet: research and policy challenges in comparative perspective, 2012). Furthermore, risk does not inevitably result to harm, but it is vital to raise awareness to the factors that give rise to the probability of harm to children.

### 4.1 Harassment and Cyberbullying

Cyberbullying is widely accepted by the academic community as a main type of risk faced by children and younger people online. A variety of psychological and social consequences have been connected to experiencing cyberbullying. These include the experience of emotional distress and anxiety, loneliness and depression, suicidal ideation and self-harm (Zych, Ortega-Ruiz, & Del Ray, 2015).

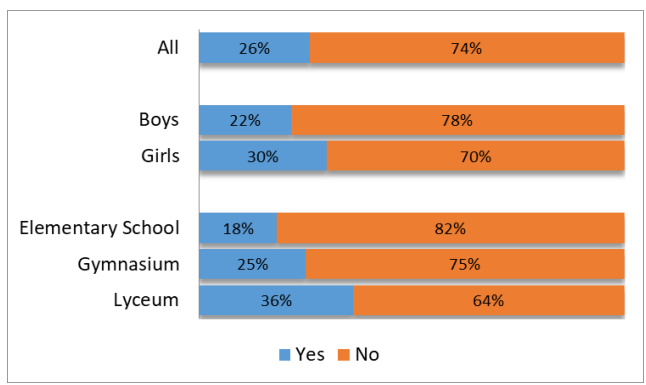
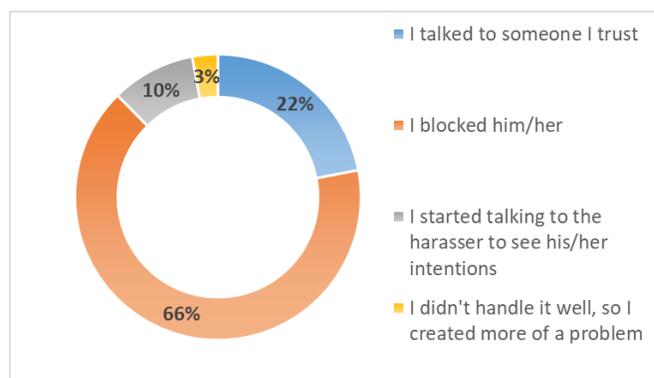

**Fig 36: Question:** Has anyone ever harassed you online?

**Fig 37: Question:** If yes, how did you react upon harassment?

According to our study, online harassment occurred to 26% of children (Fig 36). From a closer look, it appears that 18% of Elementary school, 25% of Gymnasium and 36% of Lyceum students have been at some point in their life the target of harassment. From the children that gave a positive answer, 60% are girls and 40% are boys. The majority of children who have faced online harassment (66%) reacted by blocking the harasser and only 22% talked to someone they trusted (Fig 37).

On the other hand, only 5% admits to have been victim of cyberbullying (Fig 38 up), while 17% state they have witnessed cyberbullying (Fig 38 down). Of the children who admit to have been victims of cyberbullying, 51% reported it to a trusted adult, 25% tried to ignore it, 15% dealt with it on their own, while 9% did nothing and report to continue being targeted (Fig 39 left). Of the children who admit to have witnessed cyberbullying, 21% were indifferent to the incident, 35% tried to support the victim without reporting it to an adult, while 44% responded by revealing the incident to a trusted individual (Fig 39 right).

The Greek Help-line (Help-Line, 2019) reported in their annual report of 2017 (Greek Help-Line, 2017) a high percentage of calls related to cyberbullying. Specifically, among all types of reports received by the Help-Line, cyberbullying has the highest percentage with 17% of the calls refer to cyberbullying incidences.

A research conducted by the UK Safer Internet Centre (UK Safer Internet Centre, 2017) examining the role of images in the digital lives of a representative sample of 1,500 children and young people aged 8-17 found that 22% of the sample said someone had posted an image or video to bully them. 38% reported they received negative comments on a photo they posted online. This was more frequent in older age groups (32% of 8-12 year olds compared to 45% of 13-17 year olds). Finally, 40% of the sample said that they sometimes do not post images because they are concerned about receiving negative comments.

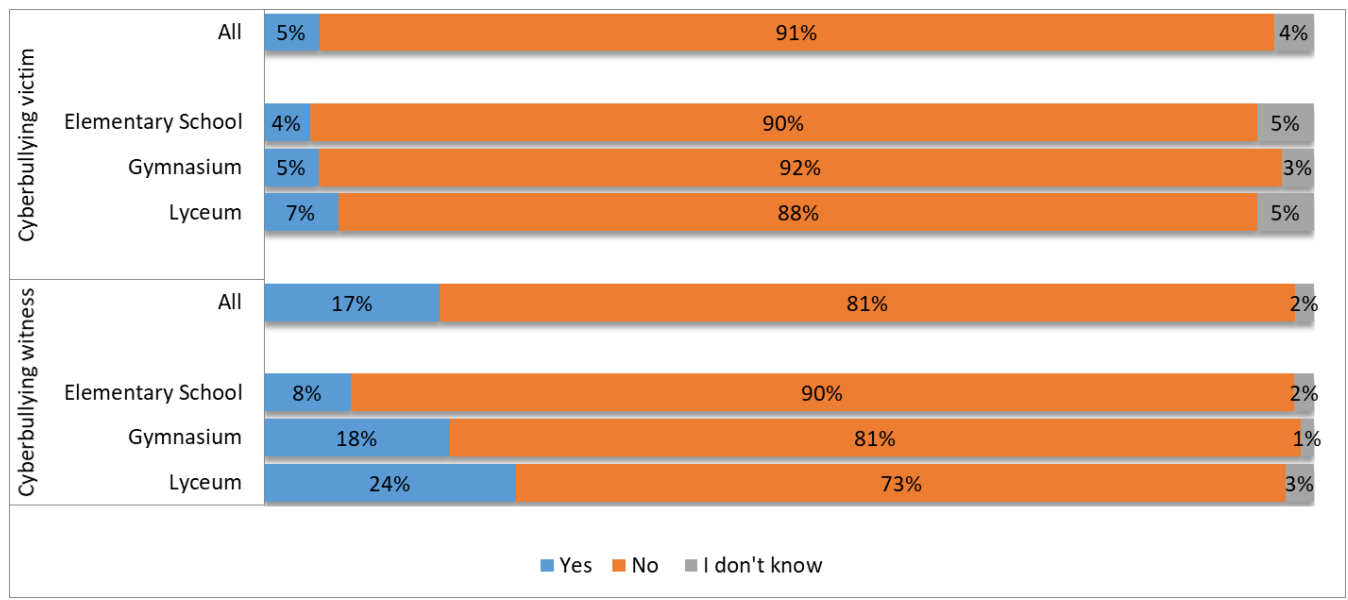

**Fig 38: Q (up):** Have you ever been a victim of cyberbullying?   **Q (down):** Have you ever witnessed cyberbullying?

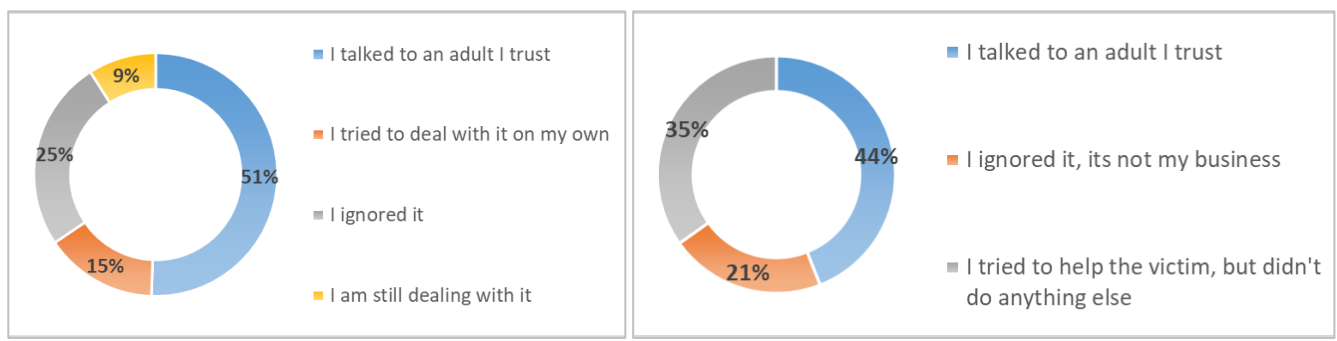

**Fig 39: Question (left):** How did you deal with cyberbullying? **Question (right):** How did you deal with witnessing Cyberbullying?

## 4.2 Meeting new people/ Strangers online

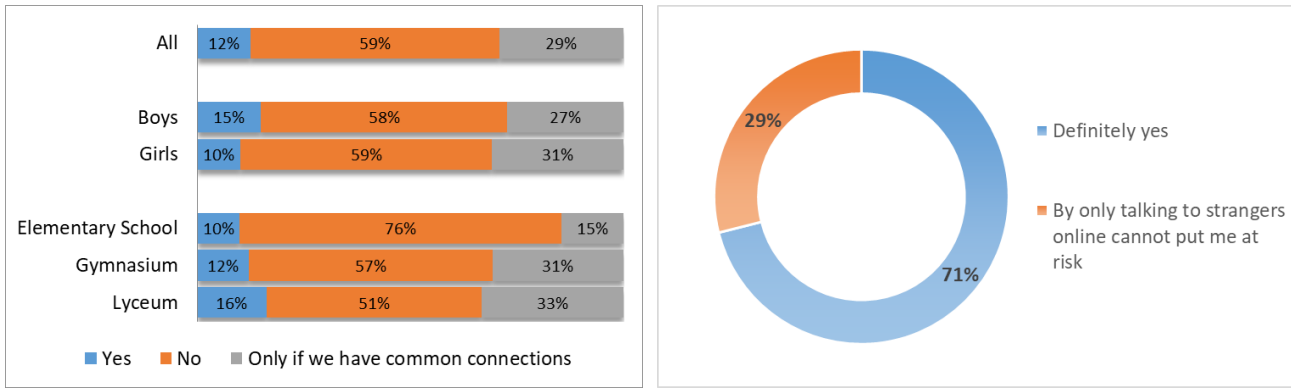

**Fig 40: Question:** Do you accept friends' requests from strangers?
**Fig 41: Question:** Do you think that talking to a stranger online can put you at risk?

An alarming percentage of 41% of children (answers: 'yes' and 'yes, if we have friend in common') accepts friend requests from strangers. 51% of them are girls and 49% are boys, thus no gender differentiation emerges. Of those that accept friend requests from strangers, 25% is from Elementary school, 43% from Gymnasium, and 49% from Lyceum (Fig 40). According to the results, 29% of them set as a condition to have common friends with the person who sent them the friend request. Therefore, children and young people consider that a friend-of-a-friend (FOAF) relationship is enough to trigger the accept button.

These results are in accordance with the percentages reported in the survey conducted by Clarke and Crowther (Clarke & Crowther, 2015). The Ofcom 2016 survey (Ofcom, 2016) reports far less percentages. This might be due to Ofcom's survey methodology, namely to take in-home interviews of the children. Our belief is that this may lead some children to be reluctant to admit that they accept friend request from strangers.

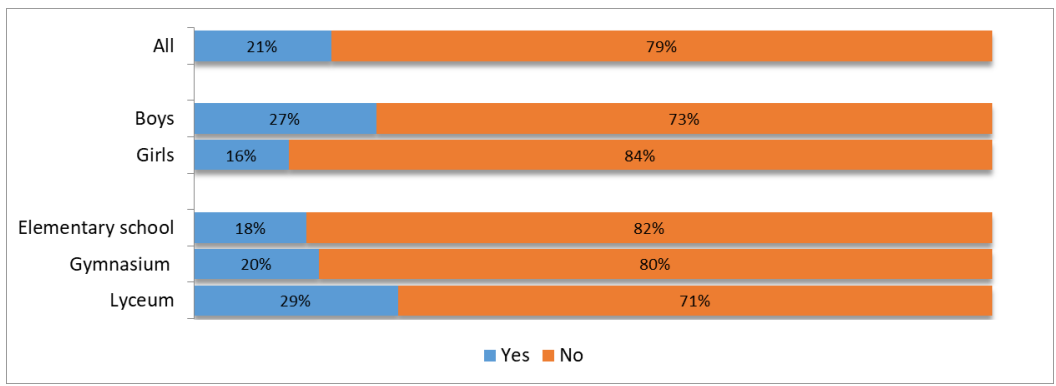

**Fig 42: Question:** Have you ever met somebody that you only knew from social media?

Moreover, our survey shows that 29% of all students do not consider that only talking to a stranger online can put themselves at risk (

Fig 41). The high percentage of 21% of children (55% boys - 45% girls) said that they have met someone they got to know online. In fact, 18% of Elementary school, 20% of Gymnasium and 29% of Lyceum students gave a positive response (Fig 42). These are admittedly high percentages showing how widespread this phenomenon is in today's children.

Stranger danger does also apply in online gaming, as children and adolescence appear to be even more open to contact strangers through an online game. In the question "Do you play online games with people you don't know in real life?" (Fig 43) 47% answers that they do play games with strangers and 53% answers that they do not play with strangers. Boys answer positively 55% while girls 32%.

In the question "Do you talk to people you don't know during an online game?" (Fig 44) 36% answer that they do talk to strangers (44% boys vs. 20% girls). It is also obvious from the response that the older the children are, the more likely it is to talk to people they don't know during an online game (29% positive answers for Elementary school students , 35% for Gymnasium students and 49% for Lyceum students). So it appears to be a common practice for children to talk to fellow players, they do not know in real word.

Finally, in the responses to the question "Have you met somebody in person that you only knew from online gaming?" (Fig 45) we observe a strange phenomenon, namely bigger percentage of younger children appear to have physically met strangers than elder children (20% of Elementary students vs. 15% of gymnasium students and 16% of Lyceum students). This might be explained from the fact that elder children appear to play less games online than younger children are (67% Elementary school students vs. 55% Lyceum students) (Fig 46).

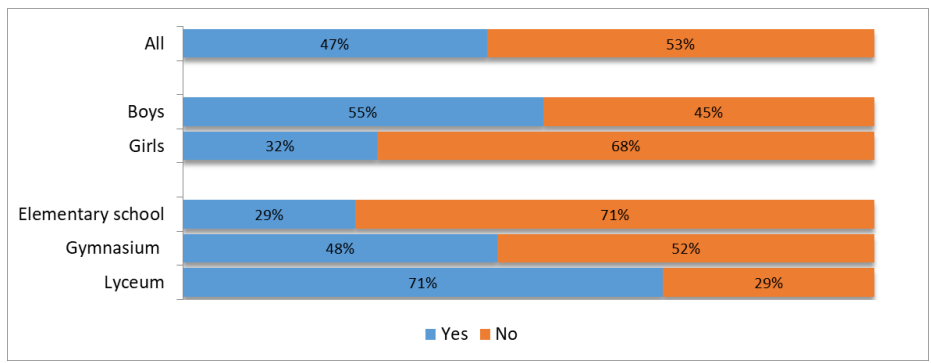

**Fig 43: Question:** Do you play online games with people you don't know in real life?

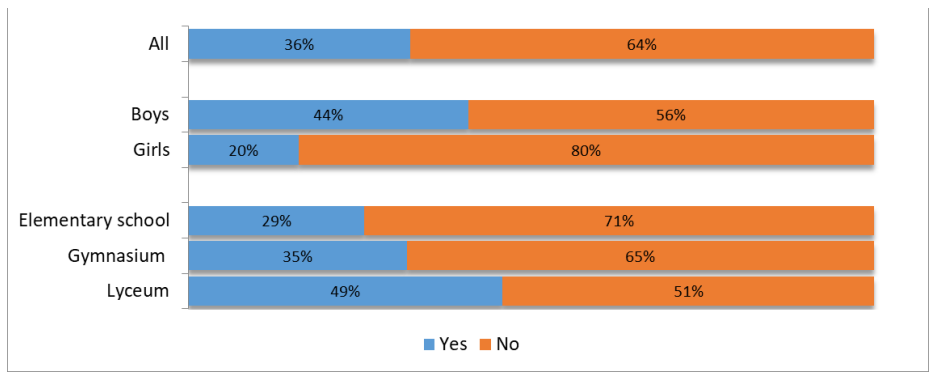

**Fig 44: Question:** Do you talk to people you don't know during an online game?

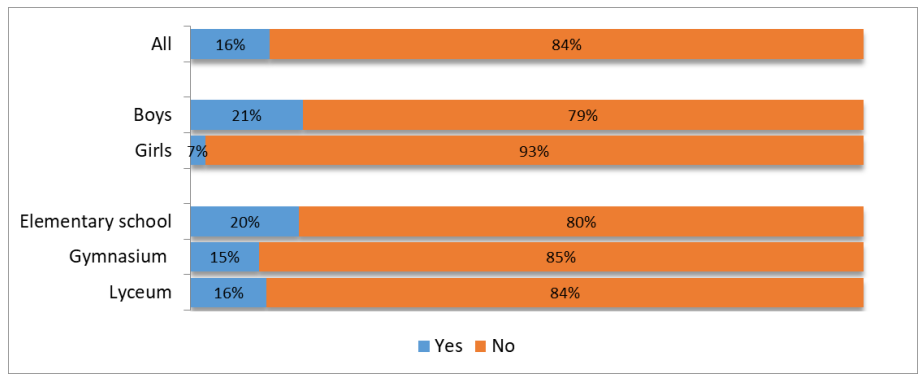

**Fig 45: Question:** Have you met somebody in person that you only knew from online gaming?

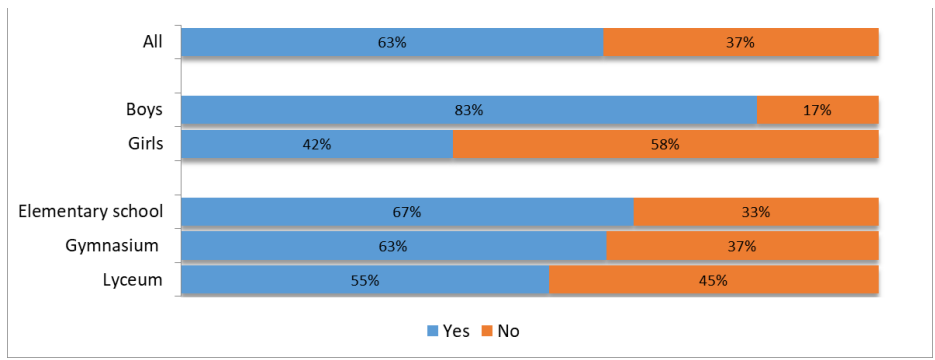

**Fig 46: Question:** Do you play games online?

The reported results that we have from the 2010 and 2014 EU Kids Online surveys (Livingstone S. , Mascheroni, Olafsson, & Haddon, 2014) about meeting online contact offline among 11-16 year old children is 9% and 13% which shows a rise in this risk. The EU Kids Online survey (Smahel, et al., 2020) reports that being in contact with someone unknown on the internet is a common experience among children (Ave = 37%). However, the prevalence of such contacts varies across countries, between 23% (Italy) and 57% (Norway). Meeting new people from the internet face-to-face is a less common experience (Ave = 16%), ranging between 5% in France to 25% in Serbia.

## 4.3 Inappropriate content

59% of students (65% boys - 54% girls) have encountered inappropriate content while surfing the internet (Fig 47). More specifically, 39% of Elementary school, 61% of Gymnasium, and 82% of Lyceum students declare to have encountered inappropriate content while online. Of all the respondents, 69% encountered it accidentally, 26% searched for it and 17% answered that it was sent from a friend (Fig 48). 24% of all children state that the inappropriate content they saw, affected them in some way. Namely, in the question "How did it affect you seeing inappropriate content" (Fig 49) it is obvious that younger children are more affected than older. Specifically, Elementary students answer 26% that they were scared, 17% that they were ashamed, and 26% that they were upset. For Gymnasium and Lyceum students the respective percentages are lower (Fig 49).

In the cross-country survey of EU Kids Online 2020 (Smahel, et al., 2020), it is reported that across the 19 countries, between 21% (France) and 50% (Serbia) of the children say they had had an experience of seeing sexual images.

On the other hand, in the 2016 Ofcom survey (Ofcom, 2016) in the question regarding experience of negative types of online/mobile phone activity among children aged 12-15 years old, only 4% has seen something of sexual nature that made them feel uncomfortable.

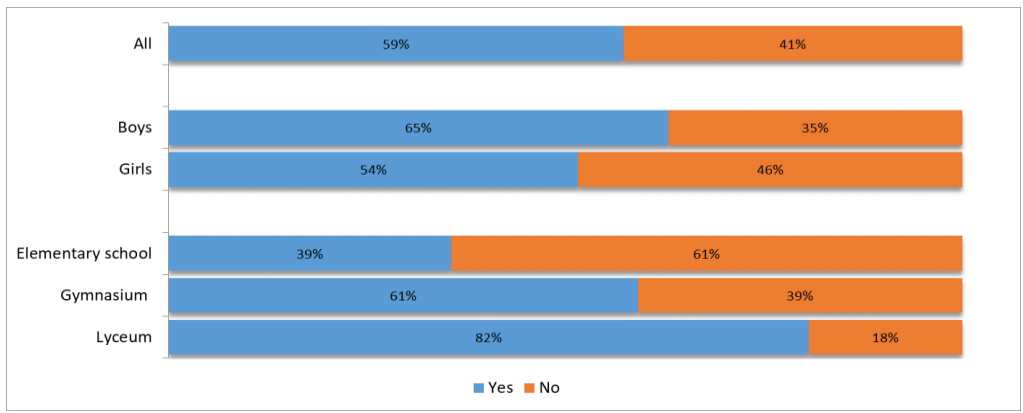

**Fig 47: Question:** Have you encountered inappropriate content online?

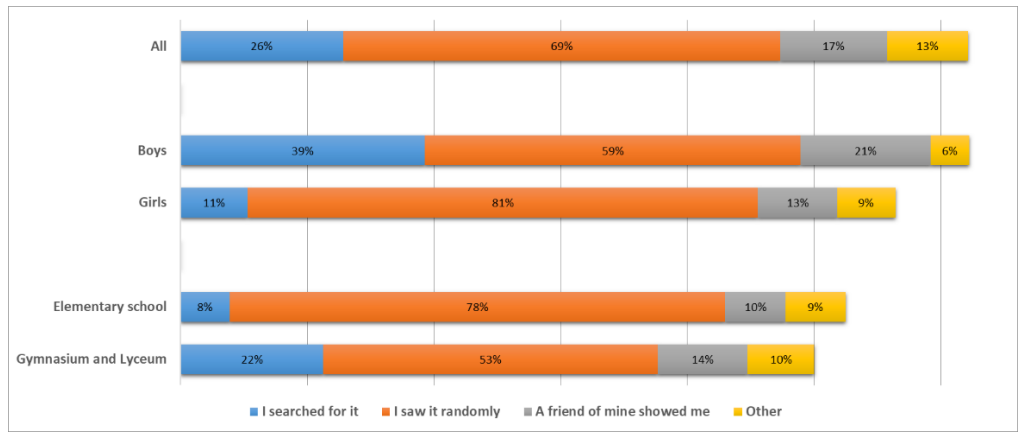

**Fig 48: Question:** Under what circumstances has it occurred?

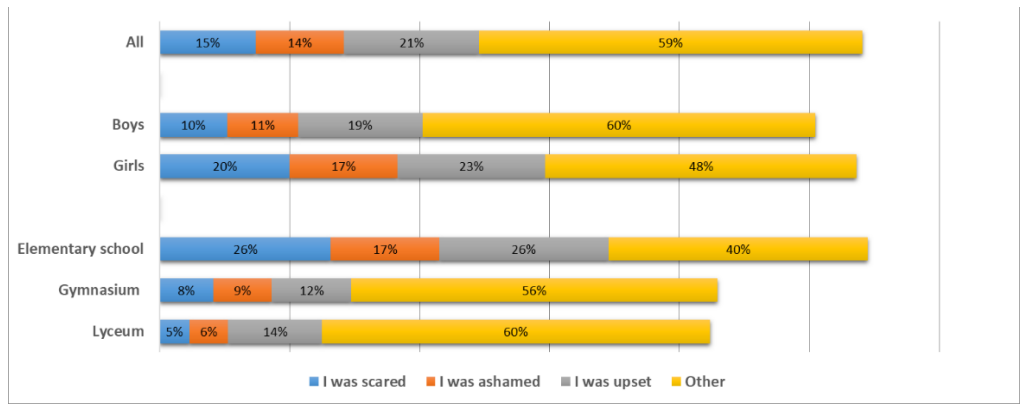

**Fig 49: Question:** How did it affect you seeing inappropriate content?

## 4.4 Exchange of personal images (sexting)

14% of children (59% boys - 41% girls) admit that they have shared very personal photos online. A positive answer came from 13% of Elementary school children, 12% of Gymnasium and 24% of Lyceum students (Fig 50). This rise in the proportion of the sample sharing very personal photos shows that the phenomenon is more common among teenagers.

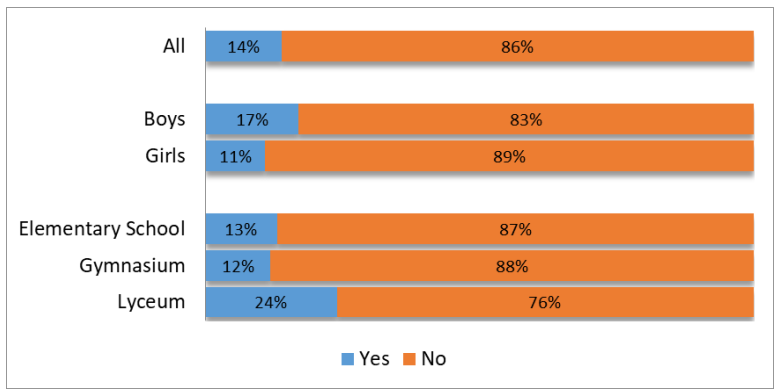

**Fig 50: Question:** Have you ever shared very personnal photos online (sexting)?

Our study shows that boys are more likely to send very personal photos. Among the children who answered that they have shared very personal photos 59% are boys and 41% are girls. UKCCIS indicates in their report (UKCCIS, 2016) that although sharing sexual images is illegal and risky, it is often the result of curiosity and exploration, and they recommend that incidents which are reported to schools should be addressed as safeguarding issue rather than criminal activity requiring involvement of the police.

Another research on the prevalence of sexting was conducted as part of a study examining the behavior in the context of romantic relationships among young people aged 14-17 (Wood, Barter, Stanley, Aghtaie, & Larkins, 2015). According to theis study, 38% of the sample had sent sexual images to a partner during or after their relationship.

According to EU Kids Online research in 19 European countries (Smahel, et al., 2020) the number of children who received sexual message in the past year ranged between 8% (Italy) to 39% (Flanders) and the average percentage between the 19 countries is 22%. More boys than girls receive sexual messages in France, Portugal, Flanders and Serbia, with the difference of 13 percentage points (Serbia) and less. On the other hand, in Finland, Germany and Lithuania, more girls report such an experience, with a difference of 13, 10 and 6 percentage points, respectivelly.

## 4.5 Excessive use of the Internet

Excessive use of the internet is a significant ongoing risk and the main focus of various studies. The Greek Safer Internet Center has attempted to record how aware children are of the amount of time they spend online and, more specifically, the time they spend on their two favorite most popular online activities, online games and social media.

Survey results reveal that 43% of children feel that they neglect other activities for the sake of the internet (answers: "very often", "often" and "yes sometimes") (Fig 51). In fact, this feeling is even stronger in Lyceum students where the rate reaches 58%. At Elementary school, 74% of children respond that they never neglect their activities for the internet.

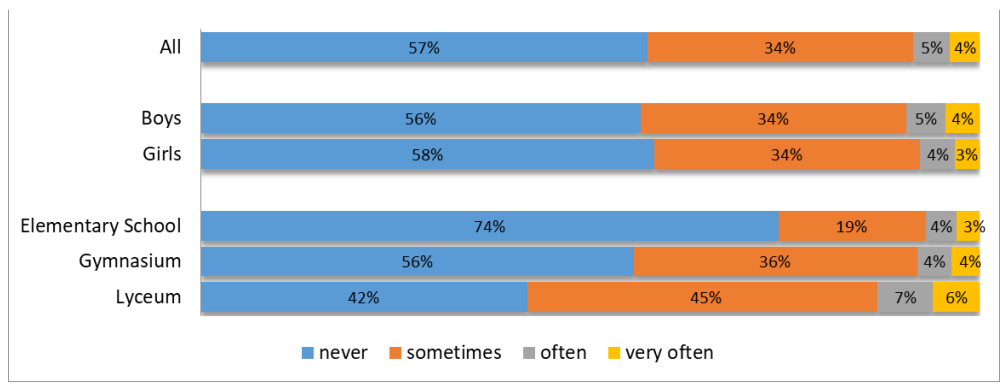

**Fig 51: Question:** Do you feel you neglect your hobbies for the internet?

Of the respondents, 19% believes that they make excessive use of the internet (answers: "yes" and "definitely yes"), while 19% of the children do not know if they use it excessively (Fig 52). Of the children who responded "definitely yes" and "yes" on the excessive use question, 52% are boys and 48% are girls.

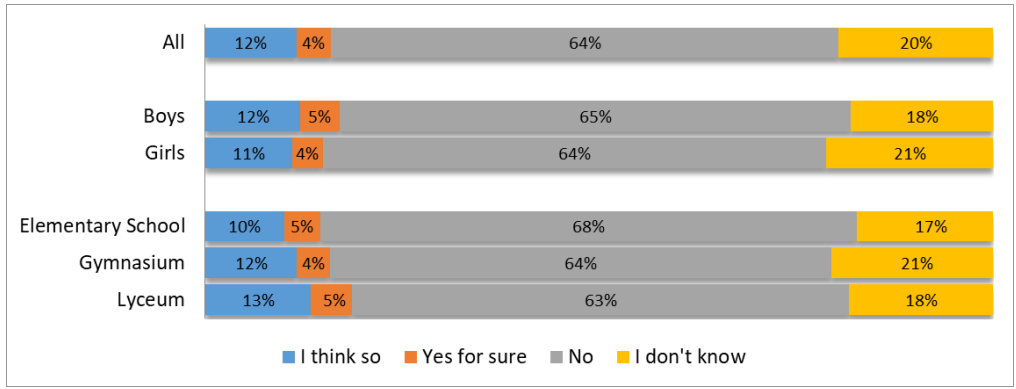

**Fig 52: Question:** Do you think you make excessive use of the internet?

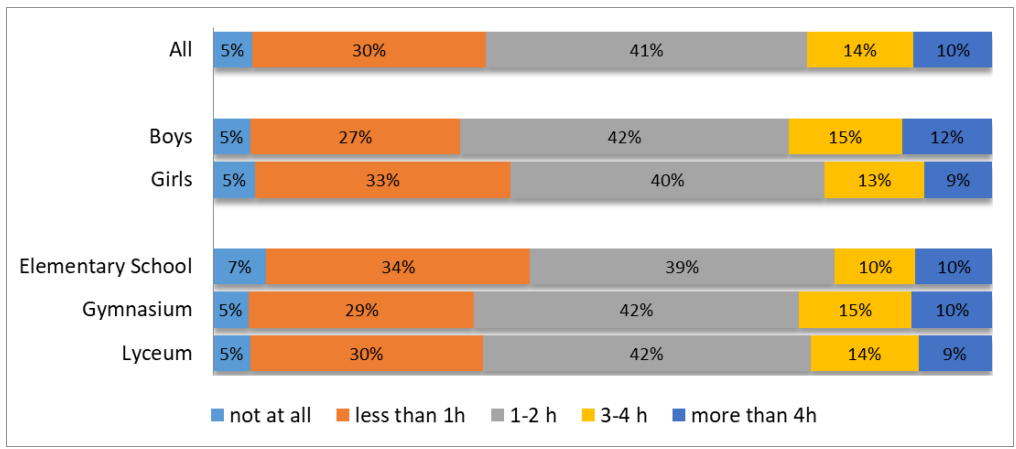

**Fig 53: Question:** How many hours per day do you spend watching YouTube videos?

71% of the children watch videos on YouTube for up to 2 hours daily, 14% of children for 3-4 hours a day and 10% of children for more than 4 hours a day (Fig 53). Percentages vary around the same levels in all ages. Of those responding more than 4 hours a day, 56% are boys and 44% are girls, so this shows a tendency for boys to spend more time on videos online.

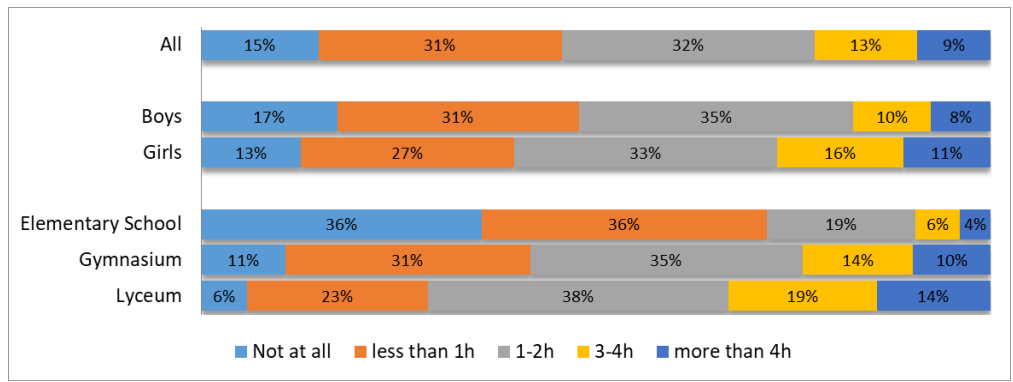

**Fig 54: Question:** How many hours do you spend on social media on weekdays?

63% of the children spend up to 2 hours daily on social media, 13% of them spend 3-4 hours and 9% more than 4 hours (Fig 54). The percentage of children who spend more than 4 hours on social networks during the weekend reaches 15%. In fact, 33% of Lyceum students spend more than 3 hours on social networks on weekdays (Fig 54) and 50% on weekends (Fig 55). For Gymnasium children the corresponding percentages are 24% for weekdays and 36% for weekends, while for Elementary school children they are 10% for weekdays and 14% for weekends, respectively.

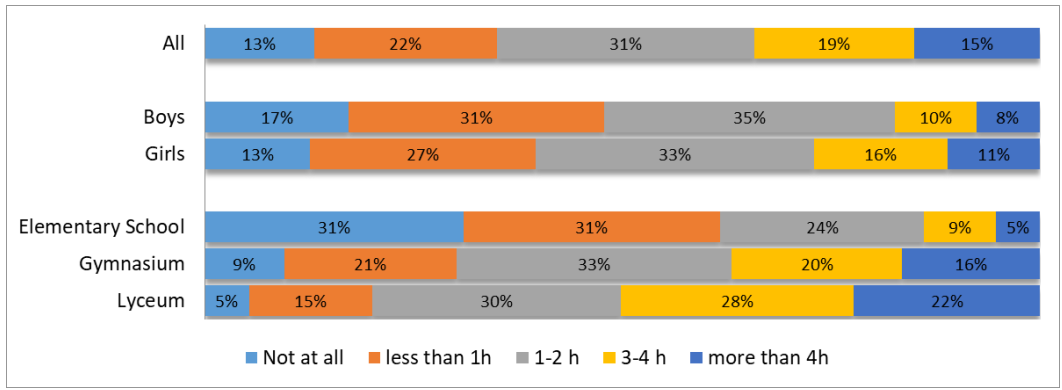

**Fig 55: Question:** How many hours do you spend on social media on weekends?

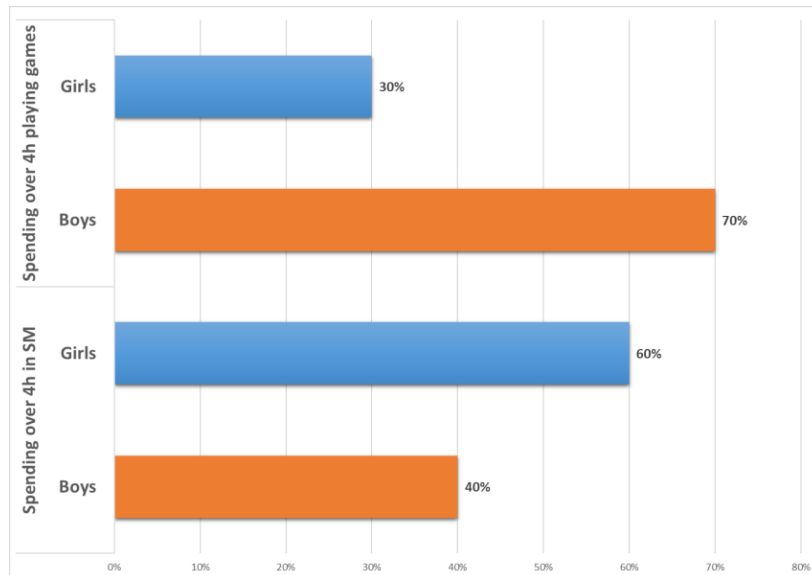

**Fig 56: (Up)** Spending over 4h per work day playing online games

**(Down)** Spending over 4h per work day in SM

It is also apparent that there is a predominance of girls (60%) over boys (40%) in the amount of time spend on social networks (Fig 56 down). This is in alignment with the study of (Livingstone, Davidson, Bryce, Batool, & Nandi, 2017) which shows that gender matters more to patterns and preferences in internet use rather than to access.

57% of the respondents play online games for up to 2 hours on a daily basis, 9% spend 3-4 hours daily while 7% spend more than 4 hours a day (Fig 57). The percentage of children who play more than 4 hours on weekends doubles and reaches 15% (Fig 58). Elementary school and Gymnasium children are mainly engaging to online games on the internet, while it is obvious that this type of activity is mainly preferred by boys (70% of boys vs. 30% of girls) (Fig 56 up).

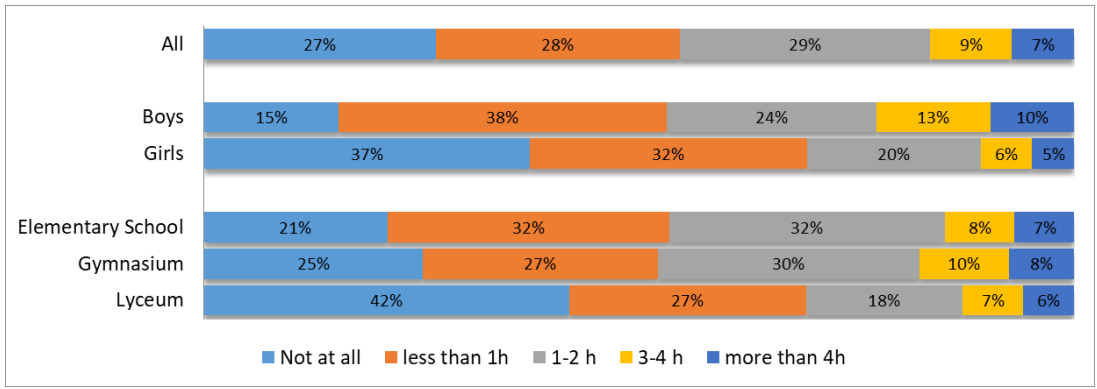

**Fig 57: Question:** How much time do you play online games on weekdays?

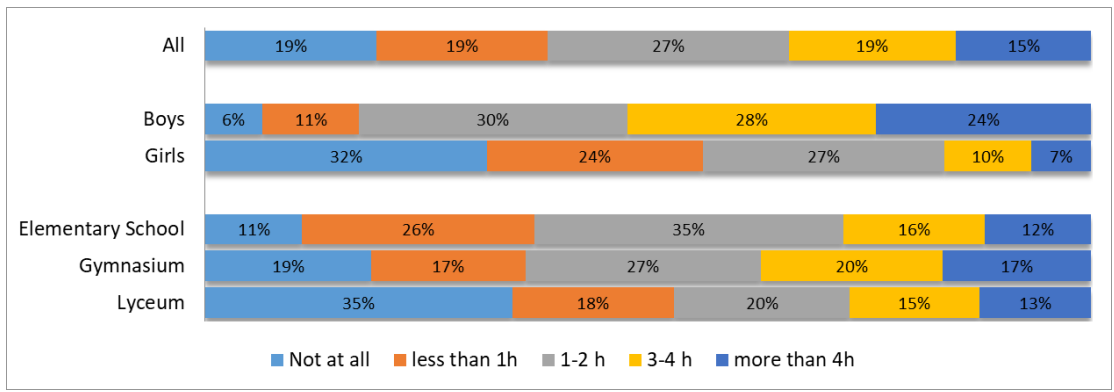

**Fig 58: Question:** How much time do you play online games on weekends?

## 5. Future Work and Conclusions

Surveys are key instruments in our work on awareness raising. They bring together the richness and depth of qualitative and quantitative research reflecting children's own voices and experiences, thus capturing empirical trends relevant to children's internet use, risks and safety. The current surveys were the first national surveys conducted in Greece concerning children's online behavior and the first that the Greek Safer Internet Center conducted where the unit of analysis were students aged 10-18 years.

Based on our survey results and analysis we can conclude with the following points; enhanced knowledge of safe internet use is important not only for children but also for parents. The Greek Safer Internet Center, through a series of awareness actions, will attempted to further sensitize the "hard-to-reach" parent target group in order to become more aware of the potential dangers hidden in inappropriate internet use, while emphasizing the benefits of the internet as a powerful medium in every aspect of life for generations to come. Furthermore, it is important to strengthen the knowledge of good practices related to internet use by children. The Greek Safer Internet Center, in collaboration with other stakeholders, will take initiatives to persuade decision makers at the Ministry of Education about the necessity of incorporating appropriate internet safety courses into school curriculums.

As emerges from the self-assessment section of the survey, the largest percentage of children feel confident that they know how to use the internet safely and how to cope with emerging risks when confronted. The survey helped emerge online risks, such as harassment and cyberbullying, accepting friends' requests from strangers, meeting with strangers, exposure to inappropriate content, sending of intimate information, etc. An important point emerging from the answers is the fact that one out of five children admits to have become the target of online harassment at some point in its 'digital life'. A significant percentage of children put themselves at risk by adopting wrong practices such as accepting friendship requests from strangers, meeting people they got to know online, sharing very personal photos on the web, sharing material without thinking of the possible consequences.

Concerning the key issue of excessive use of the internet, almost half the children admit to neglect their daily activities in order to be online, while about one out of three children believes to have an internet addiction problem. Regarding social media, an analysis of the results shows engagement of children at decreasing, not yet permissible age, most often without the consent of their parents.

Parental engagement and mediation appears to be inadequate, as involvement of parents, especially those of older children, is minimal and rules, regulations and limits are apply seldom, highlighting the need for improved parental supervision and possibly improved awareness raising and parental understanding. However, results show that parents of smaller children, the new generation of parents that also grew up with technology, is more actively involved in their child's first steps in the digital world. On the contrary, parents of older children are not sufficiently engaged to safeguard their safety, ignore the importance and fail to impose adequate rules and set limits, as they lack the required vigilance on the topic.

As future work, we intend to keep close synergy with the Ministry of Education and Religious Affairs of Greece, as well as the other Safer Internet Centers worldwide in order to conduct together research on online behavior, online risks, online time management, and social media analysis. Our intention is to include the research results in the more general Cohort study, providing insight on the way online habits and perceptions of children and adolescent change over time. Moreover, we plan to extend our survey to other target groups. We intend to create new questionnaires targeting parents where, expect from analyzing parents' and children's online habits, we'll have the opportunity to compare and contrast the beliefs of parents about their children to the answers of children themselves.

To conclude, our observations in combination with the early occupation of children with the online world, dictates the need for a more organized treatment of the topic by the educational system, starting education in schools in an organized way, incorporating material into school curricula from as early as kinder garden and preschool period.

*Internet is indeed a wonderful world, full of opportunities and challenges. Let us help our children enjoy it safely!*

## 6. Acknowledgments

This work was supported by the Greek Safer Internet Center (GA No INEA/CEF/ICT/A2018/1634563 SI4KIDS) EU project.

# 7. References

Best, P., Manktelow, R., & Taylor, B. (2014). Online communication, social media and adolescent wellbeing: A systematic narrative review. *Children and Youth Services Review*, 27 - 36.

Childwise. (2017). *Monitor Report 2017: Children Media use and purchasing.* UK: Childwise. Retrieved 7 10, 2019, from https://www.saferinternet.org.uk/research/research-highlight-series/113-childwise-monitor-report-2017

Childwise. (2019). *CHILDWISE Trends and Predictions 2019.* Norwich: CHILDWISE Research.

Clarke, B., & Crowther, K. (2015). *Children internet safety report: Key findings.* London: Family Kids and Youth, Techknowledge for schools. Retrieved 7 11, 2019, from http://www.kidsandyouth.com/pdf/FK%26Y%20Internet%20Safety%20Report%20Key%20Findings%20February%202015.pdf

Cohen, J. (1988). *Statistical Power Analysis for the Behavioral Sciences.* . New York: Lawrence Erlbaum Associates. doi: https://doi.org/10.4324/9780203771587

Coleman, J., & Hagell, A. (2008). *Adolescence, Risk and Resilience: Against the Odds.* USA: ISBN: 978-0-470-02503-1.

COUNCIL ON COMMUNICATIONS AND MEDIA. (2016). Media and Young Minds. *Pediatrics*, 10. doi:10.1542/peds.2016-2591

COUNCIL ON COMMUNICATIONS AND MEDIA. (2016). Media Use in School-Aged Children and Adolescents. *Pediatrics*, 12. doi:10.1542/peds.2016-2592

Daskalaki, E., Psaroudaki, K., & Fragopoulou, P. (2018). EL-SIC: Focus on Better and Safer Online Experiences for Kids. *ERCIM News*, 52-53. Retrieved 7 2, 2019, from https://ercim-news.ercim.eu/images/stories/EN115/EN115-web.pdf

Deming, W. E. (1990). *Sample Design in business research.* New York: John Wiley and Sons.

Directorate-General for Communication. (2019). *Special Eurobarometer 480: Europeans' attitudes towards Internet security.* Brussels: Directorate-General for Communication. Retrieved 7 1, 2019, from https://data.europa.eu/euodp/en/data/dataset/S2207_90_2_480_ENG

FORTH. (2019, 4 11). *Foundation for Research and Technology-Hellas*. Retrieved from Institute of Computer Science: www.ics.forth.gr

Global Kids Online. (2017). *Brazilian findings: more children read online news.* Sao Paolo: Regional Center for Studies on the Development of the Information Society Brazil. Retrieved 7 9, 2019, from http://globalkidsonline.net/brazilian-findings-2017/

Greek Help-Line. (2017). *annual Report of 2017.* Athens: Help-Line.gr. Retrieved 7 12, 2019, from https://saferinternet4kids.gr/ethsies-anafores/help-line-2017/

Greek Hotline for illegal Internet content. (2019, 4 11). *Greek Hotline for illegal Internet content*. Retrieved from Safeline: www.safeline.gr

Greek Safer Internet Awareness Center. (2019, 7 2). *Greek Safer Internet Awareness Center*. Retrieved from SaferInternet4kids: www.saferinternet4kids.gr

Help-Line. (2019, 4 11). *Greek Prevention Center for problematic online behaviors*. Retrieved from Help-Line: www.help-line.gr

Insafe. (2019, 4 11). *Insafe*. Retrieved from Better Internet for Kids: https://www.betterinternetforkids.eu/

Lilley, C., & Ball, R. (2013). *Younger children and social networking sites: a blind spot.* London: NSPCC.

Ling, R., & Haddon, L. (2008). *Children, youth and the mobile phone. In K. Drotner & S.Livingstone (Eds.), The international handbook of children, media and culture.* London: Sage.

Livingstone, S. (2014). Developing social media literacy: How children learn to interpret risky opportunities on social network sites. *Communications*, 283-303. doi:10.1515/commun-2014-0113

Livingstone, S. a. (2001). *Children and their changing media environment: A European comparative study.* London.

Livingstone, S., d'Haenens, L., & Hasebrink, U. (2001). Childhood in Europe:. 5.

Livingstone, S., Davidson, J., Bryce, J., Batool, S. ,., & Nandi, A. (2017). *Children's online activities, risks and safety: A literature review by the UKCCIS Evidence Group.* UK: UK Council for Internet Safety. doi: 2017

Livingstone, S., Haddon, L., & Görzig, A. (2012). *Children, risk and safety on the internet: research and policy challenges in comparative perspective.* Bristol: LSE Research Online. doi:ISBN 9781847428820

Livingstone, S., Kirwil, L., Ponte, C., & Staksrud, E. (2014). In their own words: What bothers children online? *European Journal of Communication*, 271-288. doi:https://journals.sagepub.com/doi/10.1177/0267323114521045

Livingstone, S., Mascheroni, G., & Staksrud, E. (2018). European research on children's internet use: assessing the past and anticipating the future. *New Media and Society*, 1103-1122. Retrieved 7 10, 2019, from http://eprints.lse.ac.uk/id/eprint/68516

Livingstone, S., Mascheroni, G., Olafsson, K., & Haddon, L. (2014). *Children's online risks and opportunities: Comparative findings from EU Kids Online and Net Children Go Mobile.* London: London School of Economics and Political Science.

Livingstone, S., Ólafsson, K., & Staksrud, E. (2013). Risky Social Networking Practices Among ''Underage'' Users: Lessons for Evidence-Based Policy. *Journal of Computer-Mediated Communication*, 303-320. doi:https://doi.org/10.1111/jcc4.12012

Lupiáñez-Villanueva, F., Gaskell, G., Veltri, G., Theben, A., Folkford, F., Bonatti, L., & e. a. (2016). *Study on the impact of marketing through social media, online games and mobile applications on children's behaviour.* Brussels: European Comission - Directorate-General for Justice and Consumers .

Madariaga, L., Nussbaum, M., Burq, I., Marañón, F., Salazar, D., Maldonado, L., . . . Naranjo, M. A. (2017). Online survey: A national study with school principals. *Computers in Human Behavior*, 35-44.

Mascheroni, G., & Haddon, L. (2015). Children, risks and the mobile internet. *Encyclopedia of Mobile Phone Behavior* , 1409-1418.

Mitchell, K., & Jones, L. (2012). *Youth Internet Safety Study: Methodology Report.* New Hampshire: NH: Crimes against Children.

Ofcom. (2016). *Children and parents: Media use and attitudes report.* UK: Ofcom. Retrieved 7 10, 2019, from https://www.ofcom.org.uk/__data/assets/pdf_file/0034/93976/Children-Parents-Media-Use-Attitudes-Report-2016.pdf

Ofcom. (2018). *Children and parents: Media use and attitudes report 2018.* London: Ofcom.

Park, E., & Kwon, M. (2018). Health-Related Internet Use by Children and Adolescents: Systematic Review. *Journal of Medical Internet Research*, 4. doi:10.2196/jmir.7731

Paus-Hasebrink, I., Kulterer, J., & Sinner, P. (2019). The Role of Media Within Young People's Socialisation:. *Social Inequality, Childhood and the Media. Transforming Communications – Studies in Cross-Media Research.*

Reid Chassiakos, Y., Radesky, J., Christakis, D., Moreno, M., & Cross, C. (2016). Pediatrics. *Children and Adolescents and Digital Media*, 14. doi:10.1542/peds.2016-2593

Smahel, D., Machackova, H., Mascheroni, G., Dedkova, L., Staksrud, E., Olafsson, K., . . . Hasebrink, U. (2020). *EU Kids Online 2020: Survey results from 19 countries.* London: EU Kids Online. doi:10.21953/lse.47fdeqj01ofo

Song, J. W., & Chung, K. C. (2010). Observational studies: cohort and case-control studies. *Plastic and reconstructive surgery*, 2234–2242. doi:10.1097/PRS.0b013e3181f44abc

Talves, K., & Kalmus, V. (2015). Gendered mediation of children's internet use: A keyhole for looking into changing socialization practices. *Cyberpsychology: Journal of Psychosocial Research on Cyberspace*, 9(1). doi:https://doi.org/10.5817/CP2015-1-4

UK Safer Internet Centre. (2017). *Power of image: A report into the influence of images and videos in young people's digital lives.* London: UK Safer Internet Centre. Retrieved 7 12, 2019, from www.saferinternet.org.uk/safer-internetday/2017/power-of-image-report

UKCCIS. (2016). *Sexting in schools and colleges: Responding to incidents and safeguarding young people.* London: UK Council for Child Internet Safety.

WISEKIDS. (2014). *Generation 2000: The internet and digital media habits and digital literacy of year 9 pupils (13 and 14 year olds in Wales.* Wales: WISEKIDS. Retrieved 7 10, 2019, from WISEKIDS

Wood, M., Barter, C., Stanley, N., Aghtaie, N., & Larkins, C. (2015). Images across Europe: The sending and receiving of sexual images and associations with interpersonal violence in young people's relationships. *Children and Youth Services Review*, 149-160. doi:https://doi.org/10.1016/j.childyouth.2015.11.005

Zych, I., Ortega-Ruiz, R., & Del Ray, R. (2015). Systematic review of theoretical studies on bullying and cyberbullying: Facts, knowledge, prevention, and intervention. *Aggression and Violent Behavior*, 1-21.